\documentstyle[epsfig,psfrag,color]{article}

\begin{document}

\title{The effective temperature\\
}
\author{Leticia F. Cugliandolo\\
Universit\'e Pierre et Marie Curie - Paris VI\\
Laboratoire de Physique Th\'eorique et Hautes Energies - UMR 7589
\\ 4, Place Jussieu, Tour 13, 5ft floor, 75252 Paris Cedex 05, France 
}
\date{\today}

\maketitle

\begin{abstract}
This review presents the effective temperature notion as defined from 
the deviations from the equilibrium fluctuation-dissipation theorem 
in out of equilibrium systems with slow dynamics. The thermodynamic meaning
of this quantity is discussed in detail. Analytic, numeric
and experimental measurements are surveyed. Open issues are
mentioned.

\end{abstract}

\newpage

\tableofcontents

\newpage

\textcolor{black}
{\section{Introduction}}

One of the core ideas of statistical mechanics is that
{\it equilibrium states} can be accurately described in terms of  a
small number of thermodynamic variables, such as temperature and
pressure. At present there is no equivalent framework for  
generic {\it out of equilibrium macroscopic systems}, and one is forced to 
solve their dynamics on a case-by-case basis. 

Out of equilibrium macroscopic systems are of many different kinds. 
An interesting class is the one in which the relaxation is
slow -- with observables decaying, say, as power laws instead 
of exponentially. Typical instances are coarsening phenomena
and generic glasses, realized as molecular, polymeric
or magnetic materials, among others. Another intriguing group is the one 
of non-equilibrium steady states in the weak drive 
limit. Examples are gently 
vibrated granular matter and weakly sheared super-cooled liquids.

The quest for an approximate thermodynamic description of such systems
or, to start with, the identification of effective parameters acting
as the equilibrium one, has a long history that we shall not review in
detail here. In contrast, we shall focus on the development of the
{\it effective temperature} notion, that has proven to be a successful
concept at least within certain limits that we shall discuss.

About 20 years ago, in the context of weak-turbulence, Hohenberg and
Shraiman \cite{Hosh} proposed to define an effective temperature
through the departure from the fluctuation-dissipation theorem
(FDT). However, neither a detailed analysis of this quantity nor the
conditions under which such a notion could have a thermodynamic
meaning were given in this reference. Later, Edwards mentioned the
same possibility in the context of granular
matter~\cite{Edwards}. Again, the proposal did not catch the
attention of the community. In none of the ensuing studies the
importance of distinguishing different dynamic regimes was
sufficiently stressed and the few numerical checks performed at the
time gave, consequently, confusing results. More recently, similar
ideas appeared in the glassy literature~\cite{Cukupe,Cuku3}.  In this
field, the possibility of solving exactly a number of schematic
models~\cite{Cuku1,Cuku2} put the definition of the effective
temperature, $T_{\sc eff}$, on a much firmer ground. The solution to
these models' dynamics established the importance of reaching an
asymptotic limit with slow dynamics and of attributing a value
of the effective temperature to each distinct dynamic regime. The
solutions demonstrate the relation with the phase space volume visited
dynamically~\cite{Remi}, they illustrate in many forms the
relevance of $T_{\sc eff}$ to heat transfer and equilibration, and
they set some limits to the extent of applicability of the
concept. These results opened the way to studies in a wealth of more
realistic cases. Beyond pure phenomenological descriptions of observed
phenomena used in the past that are of limited predictive power, 
see {\it e.g.}~\cite{To,fictrev,Jou}, the
$T_{\sc eff}$ definition based on fluctuation-dissipation relations is
not ambiguous and allows for direct measurements.

In this review we sift through the effective temperature notion trying
to transmit the full allure of the idea.  The literature on
fluctuation-dissipation violations is immense and several reviews have
already been
published~\cite{Crri,Calabrese-Gambassi,Corberi1,Leuzzi}. We refer to
these reports when appropriate.  The rest of the review is structured
in five Sections. The next one introduces the definition of a number
of observables and the effective temperature. It follows a discussion
of the insight gained from the solution to mean-field glassy models in
Sect.~\ref{sec:insight}. Section~\ref{sec:temperature-interpretation}
presents the interpretation of the effective temperature definition as
a {\it bona fide} temperature.  Section~5 is devoted to a
(non-exhaustive) description of numeric and experimental measurements
performed so far with special emphasis on recent studies.  Finally, in
Sect.~6 we present some brief conclusions.

\vspace{0.25cm}

\section{Definitions}
\label{sec:definitions}

\subsection{Canonical setting}

In this manuscript we focus on the dynamics of a classical or quantum
system coupled to a classical or quantum environment, typically in
equilibrium at temperature $T$.  In this sense the setting is
canonical. The systems relax by transferring energy to the environment.
Atypical dissipation, as realized in granular matter, is referred to
in Sect.~\ref{subsec:granular}. Reference to some studies in isolated 
systems (microcanonical setting) is made in the quantum context, 
see Sect.~\ref{subsec:quantum}. All along this review $k_B=1$.

\subsection{Time and length scales}

In equilibrium studies one rarely specifies the initial time or
implicitly takes it to $-\infty$. Out of equilibrium a reference moment
has to be defined. Time zero is usually taken to be the instant when
the sample is prepared in some special conditions.
In these notes we focus on infinitely rapid
quenches and we define $t=0$ as the moment when the working temperature is
instantaneously reached.

We deal mainly with macroscopic systems in which 
the number of degrees of freedom is assumed to be 
the most diverging quantity, $N\to\infty$. In 
most of the theoretical and numerical calculations 
discussed in this text times are finite with respect to $N$. 
Some reference to times that scale with the size of 
the system will be made. 

\subsection{Correlations and responses}

The out of equilibrium dynamics are explored by measuring
observables that depend on one or several times after the preparation
of the sample. Observables could be of two types: those describing
free evolution and those associated to responses to 
external perturbations. A multi-time
correlation is the average over histories or initial conditions
of several observables $O_j$, that are functions of the system's degrees of
freedom, evaluated at different times:
\begin{eqnarray}
C_{O_1O_2\dots O_n}(t_1,t_2,\dots,t_n) =\langle O_1(t_1) O_2(t_2) \dots O_n(t_n) \rangle
\; . 
\end{eqnarray}
$O_j(t)$ is any observable with, without loss of generality, zero mean 
(otherwise one simply 
substitutes $O_j\to O_j - \langle O_j\rangle$). 

In systems with an energy function the impulse response functions are
the averaged reaction of an observable to a perturbation that 
modifies the potential energy as $V\to V-h O_n$ at a given instant $t_n$. One
such response, at linear order in the perturbation, reads
\begin{equation}
R_{O_1O_2\dots O_n}(t_1,t_2,\dots,t_n) \equiv 
\left. \frac{\delta \langle O_1(t_1)\dots O_{n-1}(t_{n-1}) \rangle}{\delta h(t_n)}
\right|_{h=0}
\; . 
\end{equation}
Others, in which the perturbation(s) is (are) applied at intermediate 
time(s) can also be defined~\cite{Semerjian}.
In dynamical systems in which the time evolution of the degrees of
freedom is not related to the gradient of an energy function, one can
similarly compute the response to an instantaneous kick that translates
the degrees of freedom at a chosen instant as $x(t_n)\to x(t_n) +
\delta x(t_n)$~\cite{Vulpiani}:
\begin{equation}
R_{O_1O_2\dots O_n}(t_1,t_2,\dots,t_n) \equiv 
 \frac{\delta \langle O_1(t_1)\dots O_{n-1}(t_{n-1}) \rangle}{\delta x(t_n)}
\; . 
\label{eq:response-Vulpiani}
\end{equation}
Causality implies that both $R$'s vanish when $t_n > t_j$ with 
$j=1,\dots, n-1$. 

In cases with quantum fluctuations ones needs to distinguish the order
in which operators appear within the averages.  The symmetric and anti-symmetric
two-time correlations are
\begin{equation}
C_{[O_1,O_2]_\pm}(t_1,t_2) =
 \langle \hat O_1(t_1) \hat O_2(t_2) \rangle 
\pm 
 \langle \hat O_2(t_2) \hat O_1(t_1) \rangle 
\end{equation}
with $ \langle \hat O_i(t_i) \hat O_j(t_j) \rangle =
\mbox{Tr} [\hat O_i(t_i) \hat O_j(t_j) \hat \rho_0]/\mbox{Tr} \hat\rho_0$, $\hat \rho_0$
the initial density operator, and 
$\hat O_i(t_i)$ in the Heisenberg representation.
The linear response to $\hat H \to \hat H - h \hat O_2$ at
$t_2$ is 
\begin{equation}
R_{O_1O_2}(t_1,t_2) = 
\left. 
\frac{\delta \langle \hat O_1(t_1)\rangle}{\delta h(t_2)}\right|_{h=0}
=
C_{[O_1,O_2]_-}(t_1,t_2) \ \theta(t_1-t_2)
\end{equation}
and the last identity is the Kubo formula valid in and 
out of equilibrium at linear order in $h$.

The search for a link between dissipation and fluctuations started
with Einstein's derivation of a relation between the 
mobility and diffusion coefficient of a Brownian particle. The former is induced by an
external force, and it is a response, while the latter is due to the
spontaneous mean-square displacement of the particle's position, thus
a correlation.  Onsager's regression hypothesis, and Kubo's linear
response theory, elaborated upon the idea of an existing relation
between the two types of fluctuations near equilibrium. In the rest of
this Section we present some relations between induced and spontaneous
fluctuations that hold in and out of equilibrium.

\subsection{Fluctuation-dissipation relations}
\label{subsec:FDRs}

Several fluctuation-dissipation relations (FDRs)
between responses and correlations computed
over unperturbed trajectories have been derived in recent years. We
mention just two of these FDRs that hold in and out of equilibrium.

In generic Langevin dynamics with white~\cite{Cukupa} or
coloured~\cite{Arbicu} equilibrium noise, and Markov processes described by a
master equation with detailed balance~\cite{Corberi1}, the linear
response is related to a correlation function of the unperturbed
system which, however, is more complex than the overlap between the 
two observables involved in  the response. A simple example is
the $1d$ white-noise Langevin equation
\begin{equation}
m\ddot x(t) + \gamma \dot x(t) = F[x(t)] + \xi(t)
\; , 
\end{equation}
that using the fact that $2\gamma T R_x(t,t') = \langle x(t) \xi(t') \rangle$
implies~\cite{Cukupa} 
\begin{equation}
T R_x(t,t') = \frac12 \frac{\partial C_x(t,t')}{\partial t'} -
\frac12 \frac{\partial C_x(t,t')}{\partial t} - 
\frac{1}{2} \langle x(t) F[x(t')] - x(t') F[x(t)] \rangle 
\; . 
\label{eq:general-FDR-Langevin}
\end{equation}
$F[x]$ is any deterministic force -- not necessarily conservative.
The generalization to other observables or field-theories is
straightforward.  The extension to the non-Markovian case is more
easily performed in the path integral formalism, by exploiting the
transformation of the generating functional under
time-reversal~\cite{Arbicu}. This formulation allows one to prove that
these relation holds for Hamiltonian dynamics as well.

In the case of an Ising variable, $s=\pm $, the equivalent relation reads~\cite{Corberi1}
\begin{equation}
TR_s(t,t') =\frac12 \frac{\partial C_s(t,t')}{\partial t'} + 
\frac12 \langle s(t) \sum_{s''} [s(t') -s''] w[s(t')\to s''] \rangle
\; ,
\label{eq:general-FDR-master}
\end{equation}
with $w$ the transition probability with no perturbation applied.
This expression can be recast in the same form as
eq.~(\ref{eq:general-FDR-Langevin}) since $\langle \dot s\rangle =
\langle \sum_{s''} (s -s'') w(s\to s'') \rangle$ and the second factor
in the last term plays the role of a deterministic force. Extensions
to multi-valued discrete variables are simple.  This type of FDR can
also be generalized to intrinsically out of equilibrium systems with
transition rates that do not obey detailed balance.

Many relations of this kind have been derived in the literature. A number of 
authors re-expressed them as a sum of two contributions: the ones that
reduce to the FDT under equilibrium conditions and the anomalous ones
that allow for many intriguing interpretations (dissipated energy flux, {\it etc.}). We
refer the reader to a recent review and a few articles that summarize these ideas~\cite{Speck}
that, although certainly very interesting, are not the main focus of this article.

An FDR for dynamical systems characterized by a state variable $x$,
that reaches an invariant stationary measure, say $\rho(x)$, was
proposed by Vulpiani {\it et al.}~\cite{Vulpiani} Using the response
to an impulse perturbation of type (\ref{eq:response-Vulpiani})
one finds
\begin{equation}
R_x(t,t') =
- \langle x(t) \frac{\delta \ln \rho[x(t')]}{\delta x(t')}\rangle \
\theta(t-t')
\; . 
\label{fdr-vulpiani}
\end{equation}
Once again the generalization to a multi-variable system or a field
theory and to other observables is immediate. Another relation of this 
kind was recently proposed by several authors~\cite{Prost}:
\begin{equation}
R_x(t,t') = - \langle x(t) \frac{\delta \ln \rho_h[x(t')]}{\delta h(t')} \rangle_{h=0}
\
\theta(t-t')
\; . 
\end{equation}
The average is computed at zero field. In order to go beyond this formal 
expression one needs to know the measure $\rho_h$. Different assumptions
({\it e.g.}, Gaussian form~\cite{Vulpiani}, uniform~\cite{Martens}) lead to 
different relations. 

All these FDRs become the
fluctuation-dissipation theorem (FDT) under equilibrium conditions, as
explained in Sect.~\ref{subsec:linear}.

\subsection{Linear fluctuation-dissipation theorem}
\label{subsec:linear}

The FDT states that the decay of spontaneous fluctuations cannot be distinguished 
from the one of forced fluctuations. Its classical and quantum formulation 
are explained in the following paragraphs.

\subsubsection{Classical}

The classical linear and self fluctuation-dissipation theorem (FDT)
expresses the equilibrium thermal fluctuations of an observable $O$ in
terms of the available thermal energy, $T$, and the linear response
of $O$ to a vanishingly small applied field linearly coupled to
itself. FDT is used, for example, to
infer mechanical properties of soft matter from the fluctuations in
light scattering. Importantly enough, there is no condition on the
scale at which the observables are defined, so it could range from
the macroscopic to the microscopic.  Indeed, if one monitors the dynamics of a
classical system, either Newtonian or stochastic with detailed
balance, and assumes that equilibrium with the bath has been reached,
it follows easily that
\begin{eqnarray}
&&
R_O(t,t') 
= \frac{1}{T} \frac{\partial}{\partial t'} C_O(t,t') \
\theta(t-t')
\; .
\label{eq:FDT}
\end{eqnarray}
(This equation can be inferred from eqs.~(\ref{eq:general-FDR-master})
and (\ref{fdr-vulpiani}).  In equilibrium the last term in the
right-hand-side of eq.~(\ref{eq:general-FDR-Langevin}) vanishes due to
reciprocity and the two first ones are equal.
If $\rho$ takes the canonical form
eq.~(\ref{eq:FDT}) follows immediately from eq.~(\ref{fdr-vulpiani}).
An elegant proof for Langevin processes with generic multiplicative and 
coloured Gaussian noise is detailed in~\cite{Arbicu}.)
The integral of the response function over a time-interval running
from $t_w$ to $t$,
\begin{equation}
\chi_O(t,t_w) = \int_{t_w}^t dt' \, R_O(t,t')
\; , 
\label{eq:integrated-response}
\end{equation} 
is a dc susceptibility  that is easier to measure 
experimentally.  The FDT
implies a linear relation between $\chi_O$ and $C_O$
\begin{equation}
\chi_O(t,t_w) = \frac{1}{T} \left[ C_O(t,t) - C_O(t,t_w) \right]
\; .
\label{eq:equil-FDT-plot}
\end{equation}
A parametric plot of $\chi_O/C_O(t,t)$ against $C_O/C_O(t,t)$, constructed at 
fixed $t_w$ and for increasing $t-t_w$, is a straight line
with slope $-1/T$ joining
$(1,0)$ and $(0,\chi_{\sc eq}=1/T)$.  The same result is obtained by
keeping $t$ fixed and letting $t_w$ vary from $0$ to $t$.
A departure from the straight line
(\ref{eq:equil-FDT-plot}) for any observable $O$
signals a divergence from equilibrium.

A straightforward generalization is to the case in which the monitored observable, $A$, is not 
the same that couples to the perturbation, $B$:
\begin{eqnarray}
&&
R_{AB}(t,t') 
= \frac{1}{T} \frac{\partial}{\partial t'} C_{AB}(t,t') \
\theta(t-t')
\; .
\label{eq:FDT1}
\end{eqnarray}

In the body of this review and without loss of generality we focus on the self
correlation and response, we take $C_O(t,t)=1$, and we erase the
sub-index $O$ until discussing observable dependencies in
Sect.~\ref{sec:results}. Equation~(\ref{eq:equil-FDT-plot}) is then
more compactly written as $ \chi(C) = (1-C)/T $.

\subsubsection{Quantum}

When quantum fluctuations are important one needs to take into account 
the statistics of the observables at hand. 
Let $\phi$ and $\phi^\dagger$ be  (bosonic or fermionic)
annihilation and creation operators, respectively. 
In the Schwinger-Keldysh closed time-contour formalism 
apt to deal with real-time Green functions
these are defined  as
\begin{eqnarray}
 i\hbar G^{ab}(t,t') \equiv \langle  \phi^a(t)  \bar\phi^b(t')  \rangle
  =\mbox{Tr} \left[ 
 \mbox{T}_{\cal C} \ 
\phi_{H}(t,a) \
\phi_{H}^{\dagger}(t',b) \ \varrho_{\rm H}(0,\pm) \right]
\; ,
\end{eqnarray}
$a,b=\pm$. $\bar\phi$ is either the complex conjugate (for bosons) or the
Grassmann conjugate (for fermions) of $\phi$.
$\phi_{\rm H}(t,a)$ denotes the Heisenberg representation of the operator
$\phi$ at time $t$ on the $a$-branch of the Keldysh contour. 
$\varrho_{\rm H}(0,\pm) = \varrho(0)$ is the
initial density matrix (normalized to be of unit trace) and its location 
on the $+$ or $-$-branch is not important thanks to the cyclic property 
of the trace. In the grand-canonical ensemble
an equilibrium initial density operator reads $\varrho(0) \propto \exp\left( -\beta (H -
\mu N) \right)$, where $N$ is the number operator commuting with $H$ (in
non-relativistic quantum mechanics) and $\mu$ is the chemical potential fixing the
average number of particles.  
$\mbox{T}_{\cal C}$ is the time-ordering operator acting with
respect to the relative position of $(t,a)$ and $(t',b)$ on the Keldysh 
contour~\cite{SchwingerKeldysh,Kamenev}.
A linear transformation of the fields allows one to define Keldysh, 
retarded and advanced Green functions:
\begin{eqnarray}
&& G^K(t,t') = \frac{i\hbar}{2} [G^{+-}(t,t') +
G^{-+}(t,t')] \;, 
\nonumber \\
&& G^R(t,t')  = \left[G^{+-}(t,t')-G^{-+}(t,t')\right]\Theta(t-t')
\;, \label{eq:RelationPMBasisToLOBasis2} \\
&& G^A(t,t')  =  \left[G^{+-}(t,t')-G^{-+}(t,t')\right]\Theta(t'-t)
\;,
\nonumber
\end{eqnarray}
respectively.

The quantum fluctuation-dissipation theorem reads
\begin{eqnarray}
G^K(\omega) = \hbar
\left[\tanh\left(\frac{\beta}{2} (\hbar\omega-\mu)\right)\right]^{-\zeta} \mbox{Im} G^R(\omega)
 \end{eqnarray}
where Fourier transforms with respect to $t-t'$ have been taken and 
 $\zeta=\pm $ in the bosonic and fermionic case, respectively. 
 One can formally recover 
the canonical ensemble result by setting $\mu=0$.
 In particular, associating $G^R$ to $R$ and $G^K$ to $C$ in the notation of the 
 previous classical paragraphs and setting $\mu=0$,  in the bosonic case one has 
 \begin{eqnarray}
 R(t-t') = \frac{i}{\hbar}
\int_{-\infty}^\infty  \frac{d\omega}{\pi} \  e^{-i \omega (t-t')}  \ 
\tanh\left( \frac{\beta \hbar \omega}{2} \right) \, 
C(\omega) \, \theta(t-t')
\; , 
\label{FDTFourier}
\end{eqnarray}
that in the $\beta\hbar\to 0$ limit
reduces to the classical expression (\ref{eq:FDT1}).

\subsection{Eulerian and Lagrangian formalisms}

A class of over-damped Markov diffusion processes take an equilibrium
form -- with detailed balance -- in the Lagrangian (co-moving) frame
of the mean local velocity. The resulting stochastic process does not
contain information about the non-vanishing probability current of the
original Eulerian (laboratory) frame dynamics. In practice, though, it
is difficult to compute the local velocity and the passage to the
co-moving frame is hard to implement. Still, this observation allows
one to derive a modified FDT in the Eulerian frame that includes the
effect of the mean local velocity $\vec v$ as an additional
term~\cite{Gawedski}:
\begin{equation}
TR_{AB}(t,t') = \theta(t-t') \left[ \partial_{t'} C_{AB}(t,t') - 
\langle [ \vec v(t') \cdot \vec \nabla B(t') ] A(t) \rangle \rangle  \right]
\; . 
\end{equation}
In equilibrium the second term vanishes and the FDT is recovered. 

\subsection{Fluctuation dissipation ratio}

Consider the following adimensional object
\begin{equation}
X(t,t') \equiv
\frac{TR(t, t')}
{\partial_{t'} C(t, t')}
\; , 
\qquad\qquad\qquad
T_{\sc eff}(t,t') \equiv \frac{T}{X(t,t')}
\; , 
\label{eq:timedomain}
\end{equation}
where we assumed $t > t'$. When equilibrium is reached ${T_{\sc
    eff}}=T$. Out of equilibrium this ratio can be used to define a,
possibly two time dependent, effective temperature although the
thermodynamic meaning of this quantity is not ensured {\it a priori}.
Note the asymmetry between $t$ and $t'$ in the denominator when the
dynamics is not stationary.  Alternatively, experiments are usually
done in the frequency domain in which $ \chi(t,\omega) =\int_0^{t}
dt'' \ e^{i\omega \tau} \ R(t,t-\tau) $ and $ C(t,\omega) =\int_0^t
d\tau \ e^{i\omega\tau} \ C(t,t-\tau) $ and a time-frequency dependent
effective temperature is defined as
\begin{equation}
X(t,\omega) \equiv 
\frac{T\chi''(t,\omega)}{\omega \mbox{Re} C(\omega,t)}
\; , 
\qquad\qquad
T_{\sc eff}(t,\omega) \equiv \frac{T}{X(t,\omega)} 
\; . 
\label{freqdomain}
\end{equation}
The time-domain and frequency-domain definitions of $T_{\sc eff}$ are
not necessarily equivalent. If this quantity is to have a physical
meaning they must coincide under reasonable conditions. We shall come
back to this point in Sect.~\ref{subsec:thermodynamic}.

The asymptotic values
\begin{equation}
{X^\infty}= 
\lim_{t'\to\infty} 
\lim_{t\to\infty}
{X}(t, t') 
\; , 
\qquad\qquad\qquad
T_{\sc eff}^\infty = T/X^\infty
\end{equation}
(or $\lim_{\omega\to 0} \lim_{t\to\infty} X(t,\omega)$ 
in the frequency-domain)
turns out to be useful in the description of 
critical quenches, see Sect.~\ref{sec:critical}. 

It is clear that $X$ can be computed under any circumstances but its
interpretation as leading to an effective temperature will not always
be straightforward. In the rest of this section we shall still use the
name $T_{\sc eff}$; later in the review we shall discuss when this is
justified and when it is not.

\subsection{Beyond linear-response}

Building upon the study of non-linear responses in Langevin
processes~\cite{Semerjian}, Lippiello {\it et al.} presented a unified
derivation of out of equilibrium fluctuation-dissipation relations
(FDRs) to arbitrary order, for discrete (Ising or $q$-state) and
continuous variables ruled by stochastic Markovian dynamics of quite
generic type -- single flip, Kawasaki, Langevin, {\it
  etc.}~\cite{Lippiello} These relations are conceptually simple
although pretty lengthy to write down. They are derived from a
rewriting of the multiple variations of the evolution operator with
respect to the perturbation strength. In short, one ends up with a sum of an
increasing number of terms that are time-derivatives of higher order
correlations.

The generic FDRs can also be derived from a {\it fluctuation
principle} that simply uses the detailed balance property of the
transition probabilities~\cite{Semerjian,Arbicu,Lippiello}.  The proof
goes as follows. One first relates the probability of a path,
$\{\phi\}=\phi(t)$, under the effect of a perturbation $h(t)$ and
conditioned to the initial value $\phi(t_I)=\phi_I$ to the probability
of the reversed path, $\{\phi_R\}=\phi_R(t)=\phi(t_R)$, under the
reversed perturbation and conditioned to
$\phi_R(t_I)=\phi(t_F)=\phi_F$:
\begin{equation}
P[\{\phi\}; \phi_I, \{h\}] \ e^{-\beta \int_{t_I}^{t_F} dt \ h(t) \dot \phi(t) }
=
P[\{\phi_R\}; \phi_F, \{h_R\}] \ e^{\beta [H(\phi_I)-H(\phi_F)]}
\; . 
\end{equation}
From the average over all paths  weighted with a generic  
 distribution of initial conditions,
$P(\phi_I)$, the 
expansion in powers of $h(t)$ yields all non-linear
FDRs. This route can be followed for discrete and
continuous variables as well.

These proofs do not use any equilibrium assumption and the FDRs
hold in full generality (for aging systems, in non-equilibrium 
steady states,...).  In equilibrium the use of the Onsager
relation and the stationary property allows one to simplify the FDRs
considerably. Still, as soon as the linear regime is left, the
non-linear impulse responses are linked to sums of several
derivatives of correlations. Interestingly enough, beyond linear order the FDRs
depend on the microscopic dynamic rule not only out but also in equilibrium.

The interest in the generalized non-linear FDRs is at least
three-fold.  First, they allow one to develop efficient algorithms to
compute linear and non-linear responses without applying any
perturbation. Second, they can be used to search for growing dynamic
correlation lengths in glassy systems, a field of active
research~\cite{book-lengths}. Third, and more importantly for the purposes of this review,
they provide a way to further test the consistency of the effective
temperature notion that, in the non-linear relations between responses
and correlations, should play
essentially the same role as in the linear ones.

To our knowledge, quantum FDRs of this kind have not been derived 
yet.

\section{Insight}
\label{sec:insight}

Fully connected disordered spin models provide a mean-field
description of glassy phenomenology. They capture many features of
glassy thermodynamics and dynamics, at least in an approximate
way. They have static and dynamic transitions at finite temperatures,
$T_s$ and $T_d$, that do not necessarily coincide. In a family of such
models aimed to describe fragile glasses with the random first order
transition scenario (RFOT)~\cite{KTW1, KTW2,KTW3} $T_s$ corresponds to an entropy
crisis realizing the Kauzmann paradox.  Below $T_d$ an infinite system
does not come to equilibrium with the environment~\cite{Cuku1,Cuku2} and
relaxes out of equilibrium.  Replica tricks~\cite{Mepavi} and the
Thouless-Anderson-Palmer approach~\cite{TAP} give us a handle to
understand equilibrium and metastable states as well as thermodynamic
properties. A dynamic treatment completes the picture explaining their
relaxation and the relation with the free-energy
landscape. Importantly enough, a quasi-complete analytic solution
exists and a clear definition of many concepts that are not as sharply
grasped in finite-dimensional cases -- such as metastable states --
can be given. We do not intend to give here a full account of the behaviour
of these models; several reviews exist~\cite{PhysicaA,Cavagna} already.
We simply highlight some of their properties and their 
relevance to the effective temperature notion. 

\subsection{Slow relaxation and convergence}
\label{subsection:convergence}

The energy density as well as any
other intensive observable that depends on just one time approach
a {\it  finite} value:
\begin{equation}
 -\infty < O_\infty = \lim_{t\to\infty} \lim_{N\to\infty} \langle O(t)\rangle <+\infty
\; . 
\label{eq:asymptotic}
\end{equation}
With this order of limits, the intensive quantity $O_\infty$ is not
necessarily equal to the equilibrium value $O_{\sc eq}$. Below $T_d$ the
approach is achieved slowly, typically with a power law.
If time is let scale with $N$ 
a different  dynamic mechanism sets in and 
equilibrium  is eventually reached, $\lim_{N\to\infty} \lim_{t\to\infty} \langle O(t)\rangle 
=O_{\sc eq}$. 

\subsection{Separation of time-scales}
\label{subsection:separation}

Below their dynamic critical temperature, $T_d$, 
mean-field glassy models relax in several time-scales
 (see~\cite{Cuku2} for a precise definition). The
correlation and linear response admit a decomposition:
\begin{eqnarray}
C(t,t') &=& C^{(0)}(t,t') + C^{(1)}(t,t') + C^{(2)}(t,t') + \dots
\\
R(t,t') &=& R^{(0)}(t,t') + R^{(1)}(t,t') + R^{(2)}(t,t') + \dots
\end{eqnarray}
The first of these scales, usually labelled ${\sc st}$, remains finite as $t'
\to\infty$, while the others diverge with $t'$. In mean-field models
these time scales become infinitely separated as $t' \to\infty$ due to
their different functional dependencies on $t'$. This means that when 
one of the terms varies the other ones are either constant or have
already decayed to zero. In particular, in the ${\sc st}$ scale the correlation 
function decays from its equal-times constant value to the Edwards-Anderson 
parameter $q_{\sc ea}$ that depends on bath temperature and all coupling 
constants. In this scale the system 
follows the rules dictated by the bath and the dynamics are stationary,
while in the other scales interactions are relevant and the 
relaxation is more complex.  Each term contributing to the
correlation scales as
\begin{equation}
C^{(i)}(t,t') \simeq f^{(i)}_C\left(\frac{h^{(i)}(t)}{h^{(i)}(t')} \right)
\; ,
\end{equation}
and $h^{(0)}(t) =e^{t/\tau^{(0)}}=h^{\sc st}(t)$ with $\tau^{(0)}$ finite
and $f_C^{(0)}(\infty)=q_{\sc ea}$. Loosely, one can define a time-scale 
$\tau^{(i)}$ from
\begin{equation}
\frac{h^{(i)}(t)}{h^{(i)}(t')} \simeq 1 + \frac{t-t'}{\tau^{(i)}(t')}
\qquad\;
\mbox{with}
\qquad\;
\tau^{(i)}(t') = [d\ln h^{(i)}(t')/dt']^{-1}
\; . 
\end{equation}

One of the main ingredients in the solution to mean-field glassy
models in relaxation~\cite{Cuku1,Cuku2} is the fact that
eq.~(\ref{eq:equil-FDT-plot}) does not hold below $T_d$.  This does
not come as a surprise since the equilibrium condition under which the
FDT is proven does not apply. What really comes as a surprise is
that the modification of the spontaneous/induced
fluctuations relation takes a rather simple form:
\begin{eqnarray}
R^{(i)}(t,t') 
&=& 
\frac{1}{T^{(i)}} \frac{\partial}{\partial t'} C^{(i)}(t,t')
\ \theta(t-t') \qquad\qquad i=0, 1, \dots
\end{eqnarray}
in the $t'\to\infty$ limit.
The temperature of the external bath is the one of the 
stationary regime and the conventional FDT holds, $T^{{\sc st}}=T^{(0)}=T$. 
We discuss the reason for this in 
Sect.~\ref{sec:landscape}. In all other regimes the
$T^{(i)}$s are different from each other and from $T$. If one 
keeps $t'$ finite the crossover from one scale to another is smooth and a generic 
relation like (\ref{eq:timedomain}) applies. 

So far we highlighted three characteristics: the slow relaxation, the
fact that the energy density is bounded from below, and the separation
of time-scales. They are all fundamental for finding an effective
temperature with a thermal significance. 
 We shall come back to these properties in Sect.~\ref{sec:results}.

\subsection{The parametric construction}
\label{subsub:parametric}

A very instructive way to study the deviations from FDT
is to construct a parametric
plot of the integrated linear response against the 
correlation at fixed $t_w$, varying $t$ between $t_w$ 
and infinity. In the long $t_w$ limit such a function
converges to a limiting curve:
\begin{eqnarray}
\lim_{
\tiny{
\begin{array}{c}
t_w\to\infty\\
C(t,t_w)=C
\end{array}
}}
\chi(t,t_w)
&=&
\int_{C}^{1} dC' \  \frac{1}{T_{\sc eff}(C')}
\; , 
\label{eq:effective-T}
\end{eqnarray}
where the effective temperature~\cite{Cukupe}, $T_{\sc eff}(C)$,
is a function of the correlation $C$.  The main aims of this article
are to justify the name of this function and to give a state of the
art review of its measurements in
model systems, numerical studies and experiments (see~\cite{Crri}
for a previous survey of measurements of FDT violations with numerical 
methods and~\cite{Leuzzi} for a recent discussion on effective temperatures
from a different perspective). The convenience of this parametrization 
is that it makes it possible to compare to replica calculations and, 
on a more physical level, it allows for a better interpretation 
of fluctuations.

Let us summarize the behavior of the asymptotic $\chi(C)$ function in
different mean-field glassy models relaxing slowly with bounded
dynamics after a quench from the disordered into the ordered phases. In
all cases there is a regime $t-t_w=O(1)$ in
which $\chi(C)$ is a straight line of slope $-1/T$, joining $(1,0)$
and $(q_{\sc ea}(T),[1-q_{\sc ea}(T)]/T)$. A free-energy interpretation of this
regime will be given in Sect.~\ref{sec:landscape}. At this point the straight
line breaks and the subsequent behaviour depends on the model. Three
families have been identified.

{\it Models describing domain growth} such as, for example, the ferromagnetic 
$O(N)$ model
in $d$ dimensions in the large $N$ limit.  There is only one additional time-scale 
in which $C$ decays from $q_{\sc ea}$ to $0$. The asymptotic parametric
plot for $C \leq q_{\sc ea}(T)$ is flat. The susceptibility $\chi$
gets stuck at the value $[1-q_{\sc ea}(T)]/T$ while the correlation
$C$ continues to decrease towards zero.  The same result holds for the
Ohta-Jasnow-Kawasaki approximation to the $\lambda \phi^4$ model of
phase separation. In agreement with the fact that temperature is
basically an irrelevant parameter in (at least clean) coarsening the
effective temperature diverges in the full low-temperature phase, 
$T^{(1)}\to\infty$.

{\it Models of random-first-order-transition type} (RFOT), that provide a
mean-field-like description of fragile glasses. Examples are the
so-called $F_{p-1}$ or type B models of the mode-coupling approach and
disordered spin models with interactions among all possible $p\geq
3$-uples.  These models also have only one additional time-scale. 
In these cases the $\chi$ vs $C$ plot, for $C\leq q_{\sc
  ea}(T)$, is a straight line of negative slope larger than
$-1/T$~\cite{Cuku1}, {\it i.e.} $T<T^{(1)}<\infty$. 
The value of $T^{(1)}$ weakly depends on the
working temperature and continuously increases from $T_d$ at the
dynamic transition to a slightly higher value at $T=0$.  Another
interesting case in this group is a $d$-dimensional directed manifold
embedded in an $(N\to\infty)+d$-dimensional space under the effect of
a random potential with short-range correlations. The susceptibility-correlation
relations for different wave-vector dependent observables satisfy
$\chi_k(C_k)=\chi_{k=0}(C_{k=0})=\chi_{x=0}(C_{x=0})$ when times are
such that they all evolve in the aging regime characterized by a single 
time scale (note that $q^{\sc
  ea}_k$ depends on $k$) and, thus, the effective temperature is
$k$-independent~\cite{Cukule}.

{\it Mean-field spin models with quenched disorder having a
  genuinely continuous second-order phase transition}, for example, the
  Sherrington-Kirkpatrick spin-glass or type A models of the
  mode-coupling approach. The random manifold problem with long-range
  potential correlations is of this type too.  In these cases
  the dynamics has a continuity of time-scales, ordered in an ultrametric 
  fashion. For $C\leq q_{\sc ea}(T)$ the  
  $\chi(C)$ plot is a non-trivial curve
  with local derivative larger than $-1/T$~\cite{Cuku2}. Each value of the 
  effective temperature can be 
  ascribed to a dynamic scale.  The lowest value appears discontinuously
  ($T^{(1)} > T_{d}$) as one crosses $T_{d}$ and can be
  shown to decrease with decreasing temperature. The wave-vector
independence holds for the $T_{\sc eff}$ of the random manifold as well~\cite{Cukule,Frme}.

 Further support to the notion of effective temperatures comes from
 the study of the effect of quantum fluctuations on the same family of mean-field
 models~\cite{Culo,Malcolm}.  The setting is one in which the system is in
 contact with a quantum environment at temperature $T$ 
and the dynamics are dissipative.
 Below a critical surface (in the $T$, strength of quantum
 fluctuations, coupling to the bath phase diagram) that separates
 glassy from equilibrium phases, and in the slow dynamics regime, one
 finds a non-equilibrium relaxation with deviations from the quantum
 FDT, eq.~(\ref{FDTFourier}). These are characterized by the
 replacement of the bath temperature by an effective temperature
 $T_{\sc eff}$.  The effective temperature is again piecewise. It
 coincides with $T$ when the symmetric
 correlation $C$ is larger than $q_{\sc ea}$ and $T_{\sc eff}>T$ when
 $C$ goes below $q_{\sc ea}$.  $\chi(C)$ recovers, in the slow
 regime, the structure of the classical limit. This is
 a signal of a time-dependent decoherence effect.  The slow modes'
 $T_{\sc eff}$ depends on the working parameters and is higher than
 $T$ even at $T=0$.

Numerical simulations, with Monte Carlo
techniques or molecular dynamics of classical systems, have demonstrated
that the classification can go beyond 
the mean-field limit. We shall discuss these tests in
Sect.~\ref{sec:results}.

\subsection{Pre-asymptotic effects}
\label{subsub:pre-asymptotic}

Expression~(\ref{eq:effective-T}) is asymptotic in time, a limit in
which it is immaterial to construct the parametric plot by keeping $t_w$
fixed and letting $t$ increase to infinity, or by keeping $t$ fixed
and letting $t_w$ increase up to $t$.  However, in numerical and
experimental measurements one cannot reach this long-times limit and
it is important to have the best possible control of pre-asymptotic effects. 
The most appropriate construction at
finite times is the second option above~\cite{Sollich1}. Indeed, by
using the fact that $C(t,t_w)$ is monotonic with respect to $t_w$ one
trades all $t_w$s by $C$s and keeps $t$ as an
independent variable. Variations with respect to $t_w$ become
variations with respect to $C$ at $t$ fixed,
\begin{equation}
\frac{\partial \chi(t,t_w)}{\partial t_w} 
=
\frac{\partial \chi(t,C)}{\partial C} = - R(t,C) = 
-\frac{1}{T_{\sc eff}(t,C)} \frac{d}{dC} C = - \frac{1}{T_{\sc eff}(t,C)}
\;, 
\end{equation}
and the slope of the pre-asymptotic $\chi(t,C)$ plot at fixed $t$ yields
the searched pre-asymptotic $T_{\sc eff}$, as defined in
eq.~(\ref{eq:timedomain}).  Had we worked at fixed $t_w$ we would have
failed to obtain $T_{\sc eff}$ from the slope of the $\chi(C,t_w)$
plot. This is due to the asymmetry in the times involved in the
definition (\ref{eq:timedomain}).  The difference
disappears in the long $t_w$ and $t$ limits with $C$ varying in the
interval $[0,1]$ or for stationary systems. In cases in
which a normalization is needed, the choice of using $C(t,t)$ as the
normalization factor does not affect the slope of the plot constructed
at $t$ fixed. The FDRs discussed in Sect.~\ref{subsec:FDRs} allow one to use this method with no extra computational
effort in numerical simulations. 

\subsection{Cooling rate dependence and finite size effects}
\label{subsub:finite-size}

In the long-times dynamics of mean-field-like  models 
$T_{\sc eff}$ does not depend
permanently upon the cooling procedure. The
same models with finite number of degrees of freedom, and more refined
ones beyond the mean-field approximation, should capture a
cooling rate dependence that should  also become manifest in $T_{\sc eff}$.  

The dynamics of models in the RFOT class approach, in the asymptotic
limit $t\to\infty$ taken after $N\to\infty$, a region of phase space,
named the {\it threshold}~\cite{Cuku1}, that is higher than
equilibrium in the free-energy landscape. Further decay is not
possible in finite times with respect to $N$ since diverging barriers
separate the former from the latter. This is demonstrated by the fact
that the asymptotic values of averaged one-time intensive quantities,
such as the energy density, take higher values than in
equilibrium. Moreover, in between threshold and equilibrium a
continuous set of metastable states also separated by diverging barriers exsit.
Subsequent decay  is achieved through
activated processes in time-scales scaling
with $N$, and the relaxation of one-time quantities is expected to cross over from
power-law to logarithmic.

Although the full dynamic solution in the activated regime has not
been derived yet (it is too hard!), it is reasonable to imagine that the
$T_{\sc eff}$ values should be ordered with the higher on the
threshold and the lower, $T$, in equilibrium. The effective
temperature should relax, in logarithmic time scales, from the
threshold value to the bath temperature.  Coming back to cooling rate
dependencies, a slower cooling rate takes a finite size system below
the threshold level and the deeper the slower the rate. $T_{\sc
eff}$ should follow this variation. These claims
found some support in numerical simulations, see Sect.~\ref{sec:results}.

\subsubsection{An equilibrium interlude}
\label{subsub:equilibrium}

The replica theory of the statics of mean-field spin-glasses~\cite{Mepavi}
necessitates the definition of a symmetric disorder-averaged functional order
parameter, $P(q)$, that measures the probability
distribution of overlaps, $Nq=\sum_{i=1}^N s_i\sigma_i$, between
equilibrium configurations, $\{ s_i \}$ and $\{\sigma_i\}$. The
cumulative distribution and a further integral are
\begin{equation}
x(q) = \int_0^q dq' \ P(q')
\; , 
\label{eq:xq}
\qquad\qquad
T \aleph(C) = \int_C^1 dq \int_0^q dq' \ P(q')
\; .
\end{equation}

The out of equilibrium dynamics take place in a region of phase space
that is different from the one where equilibrium states lie. Still,
the classification of mean-field models according to $T_{\sc eff}(C)$
[or $X(C)$] and $\chi(C)$ coincides with the one arising from the
replica analysis~\cite{Mepavi} of the number and organization of
equilibrium states and their implications on $x(q)$ and $\aleph(C)$
defined in eq.~(\ref{eq:xq}). Indeed, mean-field coarsening problems
have two equilibrium states related by symmetry,
$2P(q)=\delta(|q|-q_{\sc ea})$ implying $x(q) = 0$ if $|q|<q_{\sc ea}$
and $x(q)=1$ if $|q|>q_{\sc ea}$ with $q_{\sc ea}$ the square of the
order parameter (replica symmetric -- RS -- case).  Models with a RFOT are solved by a one 
step replica symmetry breaking (1RSB)
 Ansatz implying a proliferation of equilibrium states with special properties.
Two possibilities exist for any two equilibrium configurations $\{s_i\}$ and
$\{\sigma_i\}$: they may fall in the same or the reversed state and $q=\pm
q_{\sc ea}$, respectively;  or else they may fall in different states that turn out to be
orthogonal and $q=0$. The probability distribution is hence $P(q) = x_1
\delta(q) + (1-x_1)/2 \delta(|q|-q_{\sc ea})$ and its integral yields
$x(q)=x_1$ if $|q|\leq q_{\sc ea}$ and $x(q)=1$ if $|q|>q_{\sc
  ea}$. Finally, models of the SK type are solved 
by a full RSB Ansatz, have states with all
kinds of overlaps, the $P(q)$ has two delta contributions at $\pm
q_{\sc ea}$ and a symmetric continuous part in between. This functional 
form leads to an
$x(q)$ taking the value $1$ for $|q|>q_{\sc ea}$ and a continuous
function of $q$ for $|q|$ below $q_{\sc ea}$.  In all mean-field cases
the functional forms of equilibrium and out of equilibrium objects are
similar,
\begin{equation}
x(q) \leftrightarrow X(C) 
\; , 
\qquad\qquad
 \aleph(C) \leftrightarrow 
\chi(C)
\; .
\label{eq:theorem}
\end{equation}
By this we mean that $\aleph$ has an FDT part in all cases and it is
linear, with zero, constant and finite, and variable slope, in RS, 1RSB-RFOT, and
full RSB cases, respectively.  In models of RFOT type the value of the
breaking point $q_{\sc ea}$ and the parameter $x_1$ (the slope) are
not the same dynamically and statically while in models of full RSB
kind the coincidence is complete (up to a factor $2$ due to the global
symmetry).  The same applies to all higher moments of the equilibrium
overlap $q$ and out of equilibrium $C$ distribution
functions~\cite{Cuku1,Cuku2,Frme}.  The coincidence for full RSB
models was argued to apply beyond mean-field, in finite dimensional
disordered spin systems, when the long-times limit is taken after the
thermodynamic limit.  Details of the reasoning, that is based on the
assumptions of {\it stochastic stability} and the convergence
of the out of equilibrium 
susceptibilities to the equilibrium can be found in~\cite{Frmepape}. 

These ideas were developed in spin models and one would like to 
extend them to atomic and molecular systems. However, 
overlaps
in continuous particle models are difficult to define
in a direct measurable way. Attempts based on weakly coupled real
replicas were developed in~\cite{Mepa}.  This may allow one to extend
the equilibrium $\leftrightarrow$ out of equilibrium connection to
these systems as well.

The relation between static
$\aleph(C)$ and asymptotic out of equilibrium dynamic $\chi(C)$
could 
apply in much more generality than previously suspected at the price
of identifying finite time non-equilibrium, $\chi(C,t_w)$, and finite
size equilibrium, $\aleph(C,\xi(t_w))$, with the help of a
time-dependent coherence length $\xi(t_w)$~\cite{Barrat-Berthier}.

\section{Requirements}
\label{sec:temperature-interpretation}

In this Section we list a number of conditions that the 
functional parameter $T_{\sc eff}$ must satisfy to act as 
a temperature.

\subsection{Thermometer}
\label{subsec:thermometer}

Any quantity to be defined as a non-equilibrium effective temperature
must conform to the folklore. The first requirement is to be 
{\it measurable} with a thermometer weakly and statistically coupled to
the system~\cite{Cukupe,Cuku3}.  This fact
can be proven by studying the time-evolution of the thermometer
coupled to $M$ identical copies of the system, all of age $t_w$ and evolving
independently. The thermometer is ruled by a
Langevin equation with a non-Markovian bath with statistics given by
the system's correlation and response. It thus feels the
system as a complex bath with its time scales, $\tau^{(i)}(t')$,
and temperatures, $T^{(i)}$.  The $T^{(i)}$ to be
recorded is selected by tuning the internal time-scale of the thermometer to
$\tau^{(i)}(t')$. 

Such an experiment can be relatively easily realized in systems of
particles in interaction, be them colloidal suspensions or
powders. The thermometer can be a probe particle, the free and
perturbed dynamics of which is followed in time. Diffusion is measured 
in free relaxation and mobility in the perturbed case. 
By comparing the two through an extended Einstein relation $T_{\sc eff}$ 
of the medium, that is to say, the system of
interest, is measured.  Another possibility is to monitor the kinetic energy of the
tracer (a quadratic variable) and associate it to the effective
temperature of the environment ({\it via} equipartition).  In both
cases, by playing with the tracers' parameters, namely their mass, 
different regimes of relaxation are accessed and  the $T^{(i)}$s
are measured. Consistency with the fact that the effective temperature
should be an intensive variable requires the result to be independent
of the shape of the tracers. These experiments have been performed
numerically and experimentally yielding very good
results in the former and somehow conflicting in the latter. We shall
discuss them in Sect.~\ref{sec:results}.

\subsection{R\^ole played by external baths}
\label{subsec:external-baths}

The fluctuation dissipation ratio of an `easy to equilibrate'
system should acquire the temperature of its external environment.
A rather simple though particularly illuminating problem that 
illustrates this idea is the non-Markov diffusion of a particle in a harmonic potential,
simultaneously coupled to two baths, a fast one in equilibrium at
temperature $T^{(0)}_B=T^{(0)}$ giving rise to white noise and instantaneous
friction, and a slow one with an exponential memory kernel in
equilibrium at temperature $T_B^{(1)}=T^{(1)}$.  This example can be taken
as a schematic model for an internal degree of freedom in a
slowly driven system and, in particular, the dynamics of a (possibly
confined) Brownian particle in an out of equilibrium medium. If the
slow bath is not stationary it can also be taken as a self-consistent
equation for a variable in an aging system -- exact for mean-field
disordered spin models.  The particle behaves as in equilibrium at $T^{(0)}$
or $T^{(1)}$ depending on which bath it feels. The separation is made
sharp by an adequate choice of the bath and potential energy parameters that pushes apart
their own time scales.  Similarly, an aging system with
multiple effective temperatures becomes stationary in all time-scales
with $T_B^{(i)}>T_{\sc eff}$ and goes on aging in time-scales with
$T_B^{(i)}<T_{\sc eff}$~\cite{Cuku3}.

\subsection{Observable dependencies}
\label{subsec:thermodynamic}

In cases allowing for a thermodynamic interpretation,
this intensive variable should  be the same --
partial equilibration -- for all observables evolving in the same
time-scale and interacting strongly enough.  A concrete check of 
this feature was performed within solvable glassy models. 
Two glasses in contact with a bath in equilibrium at
temperature $T$, each of them with a piecewise $T_{\sc eff}(C)$ of the
form
\begin{eqnarray*}
&& 
T^{(\sc syst \, 1,2)}_{\sc eff}(C) =
\left\{ 
\begin{array}{ll}
T & {\mbox{if}} \;\; C \geq q^{(1,2)}_{\sc ea}
\\
T^{(1,2)} & {\mbox{if}} \;\; C < q^{(1,2)}_{\sc ea}
\end{array}
\right.
\end{eqnarray*}
with $T^{(1)} \neq T^{(2)}$, were chosen. The experiment of setting two observables
in contact is reproduced by introducing a small linear coupling
between microscopic variables of the two systems. Above a critical
(though small) value of the coupling strength the systems arrange
their time-scales so as to partially come to equilibrium and the
effective temperatures below $q_{\sc ea}$ equalize. Below the critical
strength $T^{(1,2)}$ remain unaltered~\cite{Cuku3}. 

The observable independence of the fluctuation dissipation ratio 
was investigated in a
variety of out of equilibrium situations but no clear rationale as to
when this holds was found yet. An intriguing proposal was put forward
by Martens {\it et al.} who argued that the observable independence 
and hence the interpretation in terms of 
$T_{\sc eff}$ is related to the uniformity of the phase space pdf on
the hyper-surface of constant energy reached dynamically~\cite{Martens}.  These
authors validated this idea in a few simple toy models relaxing in a
single time-scale (glassy systems excluded). Consistently, this
proposal applies in mean field disordered models. Moreover, they
proposed that the observable dependence  is
proportional to the square root of the difference between the Shannon
entropy of the dynamic state and the equilibrium one, a conjecture that
deserves further investigation.

\subsection{Intuitive properties}
\label{subsubsec:intuitive}

One may wonder why in all studied un-driven systems there is a
two-time regime in which $C$ decays from its equal times value to $q_{\sc
ea}$ and FDT holds. This problem admits a physical
explanation -- thermal fluctuations within domains, rapid vibrations
within cages, {\it etc.} -- that we shall expose in
Sect.~\ref{sec:results}, a free-energy landscape
explanation that we shall discuss in Sect.~\ref{sec:landscape}
 but also a formal explanation: there exists 
a bound -- on the difference between left and right hand sides of
eq.~(\ref{eq:equil-FDT-plot}) -- that vanishes in the first regime of
relaxation~\cite{Cudeku}.

An intuitive property of $T_{\sc eff}$ is that it loosely represents
the disorder level of the system. This idea translates into $T_{\sc
  eff}$ being higher or lower than the working temperature $T$ when the initial
state is equilibrium in the disordered phase or at a 
lower temperature than the quenching value. These features
are realized in all cases studied so far, mean-field or finite
dimensional alike~\cite{Behose,Ig,Krzakala}.

This slated property matches the `fictive temperature', $T_f$, ideas that date
back to the 40s at least ~\cite{To} and have developed ever since, as explained 
in~\cite{fictrev}.  $T_f$ is
usually introduced by assuming that when a liquid falls out of
equilibrium in the glass transition region its structure gets `frozen'
at a $T_{f}$ that is higher than $T$, depends upon the cooling rate
and in particular on $T$, and deep below the transition range
approaches $T_g$. Several definitions in terms of the enthalpy or the
thermal expansion coefficient have been given  and they
do not necessarily coincide.  The fictive temperature is hence
a phenomenological convenience and 
acts essentially as a parameter in an
out-of-equilibrium `equation of state'. In contrast, the effective
temperature is defined in terms of fluctuations and responses, it
can be measured directly, and plays a r\^ole that is closer to the
thermodynamical one (although in not all possible out of equilibrium 
systems but in a class yet to be defined precisely).

\subsection{Fluctuation theorem}

The fluctuation theorem concerns the fluctuations of the entropy
production rate in the stationary non-equilibrium state of a driven
dynamical system~\cite{fluctuation-theorem}. It applies to systems
that approach equilibrium when the forcing is switched off. Of
interest in the context of glassy systems is to know the fate
of the fluctuation theorem if the system evolves out of
equilibrium even in the absence of the external drive, and whether 
$T_{\sc eff}$ enters its modified version. These questions
were addressed by Sellitto who analyzed the fluctuations of entropy
production in a kinetically constrained lattice
model~\cite{FT-previous} and by Crisanti and Ritort who discussed the
interplay between $T_{\sc eff}$ and a fluctuation theorem on
heat-exchange between the system and the
environment in the random-orthogonal model (a case in
the RFOT class)~\cite{Crri-FT}.  From a different perspective, closer to the one
in~\cite{Semerjian}, the fluctuation theorem and $T_{\sc eff}$
were studied in~\cite{Bonetto} within two 
related problems: mean-field glassy models and a
Langevin process with a number of equilibrium thermal baths with
different time-scales and temperatures as the one discussed in
Sec.~\ref{subsec:external-baths}.  Firstly, it was shown that the
work done at frequency $\omega$ by conservative and non-conservative
forces is weighted by the effective temperature (instead of the
temperature of the bath) at the same frequency.  The work of the
conservative forces produces entropy if the bath is out of equilibrium
since the nonlinear interaction couples modes at different frequency
which are at different temperature, thus producing an energy flow
between these modes.  Secondly, it was proven that the entropy
production rate satisfies a fluctuation theorem.  Thirdly, extensions of the
Green-Kubo relations for transport coefficients were 
derived. Fourthly, a feasible way to measure $T_{\sc eff}$
by exploiting the modified fluctuation theorem was discussed.
There is no satisfactory numerical test of these ideas in non-mean-field
glassy systems yet but a preliminary study will be discussed in 
Sect.~\ref{subsec:driven-numerical}.

\subsection{Non-linear effects}

In a series of papers Hayashi, Sasa {\it et al.}~\cite{Hayashi}
investigated the notion of an effective temperature in classical non-equilibrium 
steady state  (NESS)
focusing on a strongly perturbed $1d$ white-noise Langevin
system in which a Brownian particle is subject to a spatially constant
driving force $f$ and a periodic potential $U(x)$. These authors
extended the definition of $T_{\sc eff}$ that uses the
Einstein relation between diffusion coefficient and differential
mobility beyond the linear response regime, $ T_{\sc eff}(f) \equiv
D(f)/\mu(f) $. They addressed the question of the physical
significance of $T_{\sc eff}$ in at least three different ways: by
showing that it plays the role of a temperature in a large-scale
description; by proving that such an out-of-equilibrium system, used
as a thermostat for a Hamiltonian system, is able to transfer its
effective temperature as kinetic energy; and by conducting a heat
conduction experiment. The results of these tests are consistent with a 
thermodynamic interpretation.

\subsection{Local measurements}
\label{sec:local-measurements}

The fluctuations-dissipation theorem relates the averaged local response
function and fluctuations measured in equilibrium at any spatial scale. 
What should local measurements
yield out of equilibrium?

\subsubsection{Quenched disorder induced fluctuations}

In systems with quenched random interactions spatial fluctuations in
noise-averaged quantities are dictated by the local disorder.  This
fact has been known for decades; for instance, the fast and slow
character of spins in disordered magnets induced by their quenched
environment, give rise to Griffiths phenomena (free-energy
singularities, phases and slow relaxation).  In the context of glassy
dynamics and $T_{\sc eff}$, Montanari and Ricci-Tersenghi~\cite{Mori}
showed that in disordered spin models defined on random graphs spins
of two types exist: paramagnetic and glassy ones with the former
following fast equilibrium dynamics and the latter having a
non-trivial relaxation peculiar to their environment. Strictly local
deviations from FDT were characterized by an effective temperature
that can be obtained with a replica calculation along the lines
discussed in Sect.~\ref{subsub:equilibrium}.  Pretty convincing
arguments -- although not completely rigorous -- establish that in
these models the strictly local effective temperature must diffuse to
become site independent
\begin{equation}
-\frac{1}{T_{\sc eff}(C_i)} = \frac{\partial \chi_i(C_i)}{\partial C_i} 
\; . 
\end{equation}
Numerical studies of this kind of (absence of) $T_{\sc eff}$ 
fluctuations in the $3d$ EA model~\cite{Roma}
will be presented in Sect.~\ref{sec:results}.

\subsubsection{Noise induced fluctuations}
\label{subsubsec:TRI}

In a series of papers Chamon {\it et al.} proposed a theoretical
framework based on global time-reparametrization invariance that
explains the origin of dynamic fluctuations in generic -- not
necessarily quenched disordered -- glassy systems
with a separation of time-scales of the kind explained in 
Sect.~\ref{subsection:separation}~\cite{Chamon, Chamon1,Chamon2}.  Such
type of invariance had been known to exist in the out of equilibrium
dynamics of mean-field disordered models and it was later shown to
carry through to the causal asymptotic dynamics of finite $d$ infinite
size spin-glasses, under the assumption of a slow dynamics with a
separation of time-scales. The approach reviewed
in~\cite{Chamon-Cugliandolo} focuses on the fluctuations induced by
the noise at coarse-grained length scales.
The invariance acquires a physical meaning and it  
implies that one can easily change the
clock $h(t)$ characterizing the scaling of the global correlation and
linear response by applying infinitely weak perturbations that couple
to the zero mode, or with a noise-induced fluctuation. An illustration of this
property is the fact that the aging relaxation dynamics of glassy
systems is rendered stationary by a weak perturbing force that does
not derive from a potential while the $\chi(C)$ relation in the slow
regime is not much modified~\cite{Cukulepe}.  The same argument, applied to the
fluctuations, implies that easy fluctuations should be realized as
local changes in time, $t\to h_r(t)$, 
\begin{eqnarray}
\frac{h_r(t)}{h_r(t')} \simeq 1 + \frac{(t-t')}{d_{t'}\ln h_r(t')} = 1 + \frac{(t-t')}{\tau_r(t')}
\label{eq:local-ages}
\end{eqnarray}
that intervene in the local correlation and integrated linear
response.  Age measures fluctuate from point to point with younger and
older pieces coexisting at the same values of the two laboratory
times.  The two-time coarse-grained observables, {\it i.e.}
$C_r(t,t_w)$ and $\chi_r(t,t_w)$, have a slow and a fast contribution,
the former characterized by scaling functions $f_C$ and $f_\chi$ that
act as `massive variables' in the sense that they are not expected to
fluctuate in the scaling limit $\delta \ll \ell \ll \xi(t,t')$ with
$\delta$ the microscopic length-scale (lattice-spacing or
inter-particle distance), $\ell$ the coarse-graining length (to be
chosen), and $\xi$ the dynamically generated correlation length. 
Within this picture the parametric construction
$\chi_r(C_r)$ falls on the master curve for the global quantities but
could be advanced or retarded with respect to the global value with a
uniform effective temperature at fixed $C_r$ value:
\begin{eqnarray}
{T_{\sc eff}}_r(C_r)=T_{\sc eff}(C_r)
\;. 
\label{eq:local-T}
\end{eqnarray} 
This equality is non-trivial when the effective temperature is finite
and it conforms to the concept that the effective temperature for
different observables (the different regions in the sample in this
case) must equalize if they evolve in the same time-scale ({\it i.e.}
take the same value of $C_r$). Time-reparametrization
invariance is not expected to hold in cases in which $T_{\sc eff}$
diverges~\cite{Chcuyo}, notably, phase ordering kinetics (see
Sect.~\ref{subsec:coarsening}).

Equation~(\ref{eq:local-T}) with $T_{\sc eff}(C_r)$ finite implies
that the two-time variances of
composite fields, the averages of which yield the local correlation
and linear response should have the same scaling
with times~\cite{Corberi11}. These objects are easier to measure
than the joint probability distribution function (pdf) of $\chi_r$ 
and $C_r$. Numerical
results shall be discussed in Sect.~\ref{sec:results}.

\subsection{Landscapes and thermodynamics}
\label{sec:landscape}

Complex systems' dynamics are sometimes interpreted as the wandering
of a representative point in a phase space endowed with a complicated
free-energy density landscape.  The existence of an equilibrium-like
relaxation at short-time differences
suggests the distinction between `transverse' and
`longitudinal' directions in the landscape, with the former being
confining and close to, say, `harmonic' and the latter giving rise to
the slow out of equilibrium structural relaxation.  Questions
are posed as to what are these directions, which is the dynamic 
process taking the system along these directions -- diffusive, 
activated, {\it etc.} -- which is their volume in phase space, and so on.

The appearance of an effective temperature suggests some form of
ergodocity and it becomes tempting to relate $T_{\sc eff}$ to some kind
of microcanonic temperature defined from the volume of phase space
visited during the out of equilibrium excursion. The construction of a 
thermodynamics in which $T_{\sc eff}$ played a r\^ole is the natural 
step to follow. In this Section we describe the successful 
development of this program in models of the RFOT sort and how
this construction remains a suggestive picture in finite
dimensional cases.

\subsubsection{Thermal systems}
\label{subsec:thermal-syst}

Describing super-cooled liquids and glasses in terms of {\it potential
  energy landscapes} dates back to Goldstein who proposed to think of
a system's trajectory in phase space as a succession of steps among
potential energy basins~\cite{Goldstein}.  This idea developed into
the inherent structure (IS) statistical mechanics framework of
Stillinger and collaborators~\cite{Stillinger}.  In this approach each
configuration is mapped onto a local minimum of the potential energy
through a minimization process implemented, for example, by a quench
to $T=0$ (steepest descent). The inherent structure is, then, the configuration
reached asymptotically and all configurations flowing to it constitute
its basin of attraction.  Although {\it a priori} simple, this
proposal hides a number of ambiguities such as the fact that the ISs
depend on the microscopic dynamics ({\it e.g.}  single spin flip
vs. cluster spin flip in a spin system), some decision making is
needed in cases in which the $T=0$ dynamics could follow different
directions, {\it etc.} The proposal is to re-order and approximate the
partition function as a sum over IS energy levels (including their
degeneracy) times a $\beta$-dependent factor with all
contributions from the rest of the configurations -- associated to
vibrations or the fast relaxation -- basically describing the free-energy
of the liquid/glass constrained to one typical basin. The assumption
is that the IS  partition function describes the thermodynamic
properties of the state reached at very long times.

Independently, Thouless, Anderson and Palmer (TAP)~\cite{TAP} and de Dominicis and Young~\cite{deDoyo} showed that the
{\it equilibrium} properties of fully-connected spin disordered models
can be described with a local order-parameter dependent {\it
  free-energy landscape}.  Averaged observables in
equilibrium are expressed as weighted sums over the free-energies of
the TAP free-energy saddle-points:
\begin{eqnarray}
\langle O\rangle_{\sc eq} &=& 
\frac{\sum_\alpha O_\alpha \ e^{-\beta F_\alpha}}{\sum_\gamma e^{-\beta F_\gamma}}
=\frac{
\int df \ O(f) \ e^{-\beta N [f - T \Sigma(f)]}}
{\int df \ e^{-\beta N [f - T \Sigma(f)]}}
\; ,
\end{eqnarray}
in agreement with results from replica and cavity methods.  The sums
in the second member run over all stationary points of the TAP
free-energy landscape.  They transform into integrals over
free-energies at the price of introducing the number of stationary
points at given $f$, with $N\Sigma(f)=\ln {\cal
  N}(f)$ the {\it complexity}, or {\it configurational entropy} at
free-energy density $f$. This construction is {\it exact} for
mean-field models. In the $N\to\infty$ limit the integral is evaluated
by saddle-point:
\begin{equation}
\frac{1}{T} = \left. \frac{\partial \Sigma(f)}{\partial f}\right|_{f_{\sc sp}} 
\; . 
\end{equation} 
In RFOT models the disordered  state dominates above $T_d$ and
$f_{\sc eq}=f_{\sc pm}$; in between $T_d$ and $T_s$ states that are
not minima of the TAP free-energy control the integral since their
number is sufficiently large, $f_{\sc eq} = f_{\sc sp}-T\Sigma(f_{\sc
  sp})$; finally, at $T_s$ the configurational entropy vanishes
(entropy crisis) and the lowest lying, now glassy, states dominate.

The microcanonical vision of the effective temperature suggests 
to check whether below $T_d$
\begin{equation}
\frac{1}{T_{\sc eff}} = 
\left. \frac{\partial \Sigma(f)}{\partial f}  \right|_{f_{\sc th}}
\; , 
\label{eq:Teff-micro}
\end{equation}
where $T_{\sc eff}$ is the value of the slow modes effective
temperature and $f_{\sc th}$ is the free-energy at the threshold, the
level reached dynamically by an infinite system after a quench
from high $T$.  This
is indeed the case in RFOT type models in the thermodynamic limit.  Systems
with large but finite size relax
below the threshold and slowly approach equilibrium.  
Nieuwenhuizen conjectured that in these cases both $T_{\sc
  eff}$ and $\Sigma(f)$ acquire a time-dependence in such a way that
eq.~(\ref{eq:Teff-micro}) remains valid with $\Sigma(f,t)$ the
complexity of the TAP states that are relevant at time
$t$~\cite{Nieuwenhuizen}.  Moreover, in finite-dimensional systems
with short-range interactions, barrier heights and lifetimes are
finite at finite temperature and metastability becomes a matter of
time scales. A recipe to compute $\Sigma(f,t)$ in these cases was
given in~\cite{Biku}.

Once a separation of modes into fast (vibrational) and slow (structural)
is made in either an approximate (in finite dimensional models with 
short-range interactions) or exact (in mean-field cases) way, thermodynamic 
potentials that involve $T_{\sc eff}$ can be easily constructed 
and thermodynamic relations derived. This has been done in the 
IS formalism~\cite{Stillinger} and in a framework that
is closer to the TAP one~\cite{Nieuwenhuizen}.

The inadequacy of the IS approach to describe the dynamics of
coarsening and kinetically constrained models has been
explained in~\cite{Bimo}. Its limits of applicability in molecular
glasses were also discussed. Nevertheless, since it is not evident how
to access a free-energy landscape concretely, numerical efforts have
focused on the characterization of the potential energy landscape.
While increased computational facilities gave access to an exhaustive
enumeration of ISs in small clusters and proteins, the calculations
remain incomplete for macroscopic systems. A connection between $T_{\sc
  eff}$ and the IS complexity, that has to be taken with the caveats
mentioned above, was discussed in~\cite{Kob}, see 
Sect.~\ref{sec:results}.

\subsubsection{Athermal systems: Edwards ensemble}
\label{subsec:granular}

The constituents of thermal systems exchange energy with the
components of their environment and this exchange has an effect on
their motion. The constituents of {\it athermal systems} are much
larger than the ones of their surroundings and the energy received
from the bath is irrelevant.  Dissipation occurs {\it via} energy flow
from the particles to internal degrees of freedom that are excluded
from the description. Granular matter is the prototype.

In spite of the very different microscopic dynamics, the
meso/macroscopic dynamics of gently perturbed dense granular matter
share many points in common with the ones of more conventional glassy
systems. A detailed description of the experiments demonstrating these
facts is given in~\cite{Dauchot}.  

In search for a statistical mechanics description of these systems, 
 Edwards~\cite{Edwards} proposed to use the volume,
$V$, as the macroscopic conserved quantity, and the blocked
states -- defined as those in which every particle is unable to move
-- as the set of relevant equiprobable configurations,
$
P({\cal C}, V) = \Omega^{-1}(V) \theta({\cal C}) \delta({\cal V}({\cal C})- V))
$,
where $\theta({\cal C})$ is an indicator function that equals one if
the configuration is blocked and zero otherwise, ${\cal V}({\cal C})$ is the 
volume as a function of the configuration, and $\Omega(V)=\int d{\cal C}
\ P({\cal C},V)$ is the volume in configuration space occupied by the blocked
states. An entropy and compactivity are next defined as
\begin{equation}
S(V) =  -\sum_{\cal C} P({\cal C},V) \ln P({\cal C},V) = \ln \Omega(V)
\; , 
\qquad\qquad
\frac{1}{X_{\sc edw}} = \frac{\partial S(V)}{\partial V} 
\; ,
\end{equation}
respectively.  A strong hypothesis in this description is that all
blocked configurations are treated on equal footing and that any
distinction of dynamic origin is disregarded.  Moreover, the number of
conserved macroscopic variables needed to correctly describe the
system is not obvious {\it a priori}. If the grain configurations are
also characterized by an energy one should then enlarge the
description and define
\begin{eqnarray}
P({\cal C}, V,E) = \Omega^{-1}(V,E) \theta({\cal C}) 
\delta({\cal V}({\cal C})-V))  \delta(H({\cal C})-E))
\; , 
\qquad\qquad\qquad
\nonumber\\
S(V,E) =  -\sum_{\cal C} P({\cal C},V,E) \ln P({\cal C},V,E) 
\; , 
\qquad\qquad\frac{1}{T_{\sc edw}} = \frac{\partial S(V, E)}{\partial E} 
\; . 
\end{eqnarray}

The similarity between the entropy of blocked states and the
zero-tem\-perature complexity  of
glass theory is rather obvious. Indeed, the first successful check of
Edwards' hypothesis was achieved in RFOT models 
at $T\simeq 0$~\cite{Remi}. In these cases the energy is the 
relevant macroscopic variable. One  identifies
all energy minima (the blocked configurations in a gradient
descent dynamics), calculates $1/T_{\sc edw}$ and shows that it
coincides with  $T_{\sc eff}=T^{(1)}$ in the slow aging
regime ($C\leq q_{\sc ea}$)~\cite{Cuku1}.  Moreover, all intensive
observables are given by their flat average over the threshold level
$e_{\sc th} = \lim_{t\to\infty} \lim_{N\to\infty} e(t)$.   At finite $T$ the
connection can be extended at the expense of using the free-energy
instead of the energy, as explained in Sect.~\ref{subsec:thermal-syst}.

Numerical tests in finite dimensional kinetically constrained lattice
gases~\cite{Jorge-Edwards}, microscopic models of sheared granular
matter including some of the subtleties of frictional
forces~\cite{Jorge-Hernan}, spin-glasses with athermal driving between
blocked states~\cite{David-Alex} and a particle deposition
model~\cite{Brey}, gave positive results. In all these cases
Edwards measure is able to correctly reproduce the sampling of the
phase space generated by the out of equilibrium dynamics.
Nevertheless, this description does not apply to every problem with some kind of slow
dynamics.  In~\cite{Jorge-Edwards} the counter-example is the domain
growth of a $3d$ random field Ising model, a case in which
the properties of a long-time configuration of (low) energy is not
well reproduced by the typical blocked, by domain-wall pinning by
disorder, configuration of the same energy. In~\cite{Godreche-Luck},
instead, the analytically solvable one-spin flip dynamics of the $1d$
Ising chain is used to display quantitative and qualitative
discrepancies between the dynamic treatment and the averaging over an
{\it a priori} probability measure of Edwards type (and refinements).
At the mean-field level the SK model and the like do not admit a simple
relation between configurational entropy and $T_{\sc eff}$ either.

A careful account of the experimental subtleties involved in trying to put 
Edwards' hypothesis to the test, and eventually verifying whether 
$T_{\sc edw}=T_{\sc eff}$, is given in~\cite{Dauchot}. The question remains 
open especially due to the difficulty in identifying the relevant extensive 
and intensive thermodynamic parameters~\cite{Bertin-Dauchot}. 
All in all, the approach, very close to the IS and TAP constructions, 
is intriguing although not justified from first
principles yet and its limits of validity remain to be set.

\section{Measurements}
\label{sec:results}

In this Section we discuss measurements of FDT violations and tests 
of the effective temperature notion in a variety of physical systems 
out of equilibrium. Since we cannot make the description 
exhaustive we simply select a number of representative cases
that we hope will give a correct idea of the level 
of development reached in the field.

\subsection{Diffusion}
\label{subsec:diffusion}

The dynamics of a particle in a potential and subject to 
a complex environment (colored noise or baths with 
several time-scales and temperatures)  has a pedagogical
interest but also admits an experimental realization in the form 
of Brownian particles immersed in, {\it e.g.}, colloidal suspensions and 
controlled by optical tweezers. 

A particle 
coupled to a bath in equilibrium at temperature $T$ with 
noise-noise correlations of type $\langle \xi(t) \xi(t') \rangle 
\propto (t-t')^{-a-1}$, $0<a<2$, and under no external forces, 
performs normal or anomalous diffusion depending on $a$. 
The fluctuation-dissipation ratio, eq.~(\ref{eq:timedomain}), for 
$t\geq t'$ is~\cite{Pottier}
\begin{equation}
X_{xx}(t,t')= \frac{T R_{xx}(t,t')}{\partial_{t'} C_{xx}(t,t')} = 
\frac{D(t-t')}{D(t-t')+D(t')}
\; , 
\end{equation}
with the diffusion coefficient $D(t) \equiv 1/2 \ d\langle
x^2(t)\rangle/dt \simeq t^a$ for $a \neq 1$ and $D(t) = \mbox{ct}$ for
$a=1$. In the colored noise cases $X_{xx}$ is a non-trivial function of
times and it does not seem to admit a thermodynamic
interpretation. Still, for later reference we
consider the long times limit:
\begin{eqnarray}
\lim_{t'\to\infty} \lim_{t\to\infty} 
X^\infty = X_{xx}(t,t') \simeq \left\{
\begin{array}{ll}
0  \qquad & a< 1 \qquad \mbox{subOhmic,}
\nonumber\\
1/2 \qquad & a=1 \qquad \mbox{Ohmic,}
\label{eq:diffusion}\\
1 & a>1 \qquad \mbox{superOhmic.}
\end{array}
\right.
\end{eqnarray}

Another illustrative example is the non-Markovian diffusion of a
particle in a harmonic potential and subject to different external
baths~\cite{Cuku3,Bonetto,Ilg2}. As already explained in
Sect.~\ref{subsec:external-baths} this simple system allows one to
show how different environments can impose their temperatures on
different dynamic regimes felt by the particle.  Tests of other
definitions of out of equilibrium temperatures in this simple case
confirmed that the definition that appears to have the most sensible
behaviour is the one stemming from the long-time limit of the
relations between induced and spontaneous
fluctuations~\cite{Ilg2}. All other definitions yield results that are
more difficult to rationalize: in most cases one simply finds the
temperature of the fast bath and in some cases, as with the static
limit in~\cite{OHern}, one incorrectly mixes different time regimes
even when their time-scales are well separated.

\subsection{Coarsening}
\label{subsec:coarsening}

When a system is taken across a second order phase transition into an
ordered phase with, say, two equilibrium states related by symmetry,
it tends to order locally in each of the two but, globally, it remains
disordered. As time elapses the ordered regions grow and the system
reaches a scaling regime in which time-dependencies enter only through
a typical growing length, $L(t)$.  Finite dimensional coarsening
systems have been studied in great detail from the effective
temperature perspective (see~\cite{Corberi1}). In this context, it is
imperative to distinguish cases with a finite temperature phase
transition and spontaneous symmetry breaking from those with ordered
equilibrium at $T=0$ only.  Some representative examples of the former
are the clean or dirty $2d$ Ising model
with conserved and non-conserved order parameter.  An instance of the
latter is the Glauber Ising chain and we postpone its discussion to
Sect.~\ref{subsec:chain}.

Let us focus on scalar systems with discrete broken symmetry.
When time-differences are short with respect to the typical growing
length $L(t_w)$, domain walls remain basically static and the only
variation is due to thermal fluctuations on the walls and, more
importantly, within the domains. This regime is stationary, and induced
and spontaneous fluctuations are linked by the FDT. At longer time-differences
domain walls move and observables display the out of equilibrium 
character of the system.

The correlation and total susceptibility in the $t_w\to\infty$ limit
separate in two contributions 
$C(t,t_w)= C^{\sc st}(t-t_w) + C^{(1)}(t,t_w) $ and 
$\chi(t,t_w) = \chi^{\sc st}(t-t_w) + \chi^{(1)}(t,t_w)$.  Numerical studies of $T_{\sc eff}$ 
focused on the
parametric construction $\chi(C,t_w)$ at fixed and finite $t_w$ where
the chosen observable is the spin itself. The resulting plot has a
linear piece with slope $-1/T$, as in eq.~(\ref{eq:equil-FDT-plot}),
that goes below $C=q_{\sc ea}=m^2$ and, consistently, beyond
$\chi=[1-m^2]/T$.  The additional equilibrium contribution is due to
the equilibrium response of the domain walls that exist with finite
density at any finite $t_w$.  In the truly asymptotic limit their
density vanishes and their contribution disappears. Consequently,
$\lim_{t_w\to\infty}\chi(C,t_w)=C^{\sc st}\geq q_{\sc ea}$ satisfies
FDT and it is entirely due to fluctuations within the domains. In cases
with $L(t)\simeq t^{1/z_d}$, the slow
terms take the scaling forms
\begin{equation}
C^{(1)}(t,t_w) \simeq f_C(t/t_w)
\; , 
\qquad\qquad
\chi^{(1)} (t,t_w) \simeq t_w^{-a_\chi} \ f_\chi(t/t_w)
\; . 
\end{equation}
It would be natural to assume that $\chi^{(1)}(t, t_w)$ is
proportional to the density of defects $\rho(t) \simeq L(t)^{-n}
\simeq t^{-n/z_d}$ with $n = 1$ for scalar and $n = 2$ for vector
order parameter.  Although this seems plausible
$a_\chi$ is instead $d$-dependent. Another conjecture is~\cite{Corberi1}
\begin{eqnarray}
z_d \ a_\chi =
\left\{
\begin{array}{ll}
n \  (d-d_L)/(d_U-d_L) & \qquad \qquad d<d_U \; , 
\\
n \;\;\;\;  (\mbox{with} \; \ln \; \mbox{corrections}) & \qquad \qquad d=d_U \; , 
\label{eq:exponent-achi}
\\
n  & \qquad \qquad d>d_U \; . 
\end{array}
\right.
\end{eqnarray}
$d_L$ is the dimension at which $a_\chi$ vanishes and may coincide
with the lower critical dimension.  One finds $d_L=1$ in the Ising
model, $d_L=1$ in the Gaussian approximation of Ohta, Jasnow and
Kawasaki, and $d_L=2$ in the $O(N)$ model in the large $N$
limit. $d_U$ is the dimension at which $a_\chi$ becomes
$d$-independent and it does not necessarily coincide with the upper
critical dimension. One finds $d_U=3$ in the Ising model, $d_U=2$ in
the Gaussian approximation, and $d_U=4$ in the large $N$ $O(N)$ model.
 It was
then suggested that $d_U$ might be the highest $d$ at
which interfaces roughen.  In all cases in which $a_\chi>0$, $T_{\sc
eff}\to\infty$.  This result was confirmed by studies of second order
FDRs in the $2d$ Ising model that showed the existence
of stationary contributions verifying the non-linear equilibrium
relation and aging terms that satisfy scaling and yield $T_{\sc
eff}\to\infty$ as in the linear case~\cite{Lippiello}.  The approach
by Henkel {\it et al.} based on the conjecture that the response
function transforms covariantly under the group of local scale
transformations, fixes the form of the scaling function $f_\chi$ but
not the exponent $a_\chi$~\cite{Henkel} and does not make
predictions on $T_{\sc eff}$. The coincidence between statics and dynamics,
see Sect.~\ref{subsub:equilibrium}, holds in these cases~\cite{Corberi1}.

Noise induced spatial fluctuations in the effective temperature of
clean coarsening systems were analyzed in the large $N$ $O(N)$
model~\cite{Chcuyo} and with numerical simulations~\cite{Cocu}. The
first study shows that time-reparametrization invariance is not
realized and that $T_{\sc eff}$ is trivially non-fluctuating in this
quasi-quadratic model. The second analysis presents a conjecture on
the behaviour of the average over local (coarse-grained)
susceptibility at fixed local (coarse-grained) correlation that
consistently vanishes in coarsening (but is more interesting in
critical dynamics as we shall discuss in Sect.~\ref{sec:critical}).
 
The results gathered so far and summarized in the conjecture 
(\ref{eq:exponent-achi}) imply that the FD ratio vanishes
and thus $T_{\sc eff}$ diverges in quenches into the ordered 
phase of systems above their lower critical dimension. This

\subsection{Critical dynamics}
\label{sec:critical}

The non-equilibrium dynamics following a quench from the disordered state 
to the critical point
consists in the growth of the dynamical correlation length, $\xi(t)
\simeq t^{1/z_{eq}}$. This length does not characterize the size of
well defined domains but the size of a self-similar structure of
domains within domains, typical of equilibrium at the critical point. 
A continuum of {\it finite} time-scales
associated to different wave-vectors, $\tau^{(k)} \simeq k^{-z_{eq}}$,
exists with only the $k\to 0$ diverging.  At any finite 
time $t$, critical
fluctuations of large wave-vectors, $k\xi(t) \gg 1$, are in almost
equilibrium, while those with small wave-vectors, $k\xi(t) \ll 1$,
retain the non-equilibrium character of the initial condition.  This
{\it finite-time}
separation, and the fact that the order parameter vanishes, leads to
the {\it multiplicative} scaling forms
\begin{eqnarray*}
&& 
C(t, t_w) \simeq \xi(t-t_w)^{-d+2-\eta} \ f_C[\xi(t)/\xi(t_w),\xi_0/\xi(t_w)]
\; , 
\\
&&
\chi(t, t_w) \simeq \beta - \xi(t-t_w)^{-d+2-\eta} \ f_\chi[\xi(t)/\xi(t_w),\xi_0/\xi(t_w)]
\; ,
\end{eqnarray*}
with the microscopic length $\xi_0$ ensuring the normalization of the 
correlation and the fact that $\chi$ vanishes at equal times.
These forms imply that beyond the initial equilibrium part, the
$\chi(C)$ plot assumes a non-trivial shape that, however,
progressively disappears and approaches the equilibrium linear form at
all $C>0$. The limit $C=0$ is distinct and 
the limiting parameter $X^\infty$ should be non-trivial and universal
in the sense of the renormalization
group~\cite{Godreche-Luck-critical}. Whether this one can be interpreted
as a temperature is a different issue that has been only partially 
discussed. For this reason, we keep the notation $X^\infty$ 
(instead of $T_{\sc eff}$) in most of this section. 

The correct estimation of $X^\infty$ has to take into account that the
number of out of equilibrium modes decreases in the course of time
(contrary to what happens in the random manifold problem in the large
$N$ limit, for example).  The best determination of $X^\infty$ is
achieved by selecting the $k\to 0$ mode.  A thorough review of the
universality properties of $X^\infty$ found with the perturbative
field-theoretical approach and some exact solutions to simple models,
as well as the comparison to numerical estimates, is given
in~\cite{Calabrese-Gambassi}.  In conclusion, $X^\infty$ is a
universal quantity that does not depend on the observable -- as
checked for a large family of them -- but recalls certain features of
the initial condition~\cite{Krzakala} and the correlations of the
environment~\cite{Julius}. In the scalar model one finds the diffusive
results, eq.~(\ref{eq:diffusion}), at the Gaussian level and
corrections when higher orders are taken into account. For example
$X^\infty=0.30(5)$ in $d=2$, $X^\infty=0.429(6)$ in $d=3$ for a quench from 
a disordered state, white noise and up to second order in $4-d$.
Instead, $X^\infty\simeq 0.78$ ($d=3$) and $X^\infty=0.75$ ($d=2$)
if the initial state is magnetized. A larger $X^\infty$ implies a
lower $T^\infty_{\sc eff}=T_c/X^\infty$ and the comparison 
between these values conforms to the intuitive idea that 
an ordered initial state leads to a lower effective temperature 
than a disordered one. 

A different type of critical phenomena (infinite order) arises
in the $2d$ XY model.  The magnetic order parameter vanishes at all
$T$ but there is a low-$T$ critical phase with quasi long-range order
(power-law decaying spatial correlations) that is destroyed at
$T_{\sc kt}$ where vortices proliferate and restore a finite
correlation length. Out of equilibrium the critical scaling forms
 apply although with a temperature-dependent exponent, $\eta(T)$, 
 and a growing length
scale $\xi(t)\simeq (t/\ln t)^{1/2}$ (the logarithm is the effect of vortices). 
The r\^ole of
the EA order parameter is played by the asymptotically vanishing
function $(t_w/\ln t_w)^{-\eta(T)/2}$ and the crossover between equilibrium and
out of equilibrium regimes takes place at a $t_w$-dependent value of
the correlation.  The $\chi(C,t_w)$ plot at finite $t_w$ is curved, it
does not reach a non-trivial master curve for $t_w\to\infty$, but
$T_{\sc eff}(t,t_w)=f_X[\xi(t)/\xi(t_w)]$.  Quenches from
the disordered phase, $T_0>T_{\sc kt}$ and heating from a $T_0=0$
ground state to $T<T_{\sc kt}$ demonstrate that the slow modes' $T_{\sc
  eff}$ depends on the initial state and it is higher (lower) when
$T_0>T$ ($T_0<T$)~\cite{Behose}. We allow ourselves to use $T_{\sc eff}$ in this 
case since these results point in the direction of justifying its thermodynamic
meaning. Similar results were obtained for
$1+1$ elastic manifolds with and without quenched disorder
(see~\cite{Ig} and refs. therein). As the dynamic-static link is
concerned, Berthier {\it et al.} evinced that the extension to
finite-times finite-sizes works, at least at not too high $T$s where
free vortices inherited from the initial condition are still present.
The coexistence of a single time scale in the aging regime together
with a smooth and time-dependent $\chi(C,t_w)$ plot arises naturally in a
critical regime and it is due to the lack of sharp time-scale
separation.

The exact calculation of the joint probability distribution of the
finite-size correlation and linear response in the spherical
ferromagnet quenched to its critical temperature was given 
in~\cite{Annibale}.  The results prove that these fluctuations are not
linked in a manner akin to the relation between the averaged
quantities, as proposed in~\cite{Chamon-Cugliandolo}, see 
Sect.~\ref{subsubsec:TRI}, 
for glassy dynamics. The correlation-susceptibility
fluctuations in non-disordered finite-dimensional 
ferromagnets quenched to the critical
point were examined in~\cite{Cocu} where it was shown that the
restricted average of the susceptibility, at fixed value of the
two-time overlap between system configurations, obeys a scaling
form. Within the numerical accuracy the slope of the scaling function
yields, in the asymptotic limit of mostly separated times, the
universal value $X^\infty$.

The first experiments testing fluctuation dissipation deviations in a liquid
crystal quenched to its critical point appeared recently and the
results are fully consistent with what has been discussed
above~\cite{Joubaud}.

Although many evaluations of $X^\infty$ in a myriad of models tend to
confirm that it mostly behaves as a critical
property~\cite{Calabrese-Gambassi}, the thermodynamic nature of this
parameter has not been explored in full extent yet. Measurements with
thermometers and connections to microcanonical definitions have not
been performed at critical points.

\subsection{Quenches to the lower critical dimension}
\label{subsec:chain}

The kinetic Glauber-Ising spin chain is the protoype of a dynamic
model at its lower critical dimension.  Taking advantage of the fact
that this is one of the very few exactly solvable models of
non-equilibrium statistical mechanics, several issues concerning the
effective temperature interpretation have been addressed in this case,
as the observable dependencies.

After a quench from $T_0\to\infty$ to $T=0$
the factor $X_s(t, t')$, associated to the 
spin correlation and susceptibility, 
 is $\leq 1$ and its value $X_s^\infty$ in the limit $C_s
\to 0$ evolves smoothly from $1/2$ (as in models characterized by
simple diffusion such as the random walk or the Gaussian
model~\cite{Cukupa}) to $1$ (equilibrium) as $t/\tau_{eq}$ grows from
$0$ to $\infty$ [$1/\tau_{eq}=1-\tanh(2J/T)$ is the
smallest eigenvalue of the master equation operator].  Moreover,
$X_{s}$ is an exclusive function of the auto-correlation $C_{s}$
as in more complex instances of glassy behaviour~\cite{1dIC}.

The value for the long-wavelength analogue, 
the fluctuating magnetization, $X_m^\infty$, is 
identical to the local value $X_s$. The physical origin of the
local-global correspondence, which can also be obtained by
field-theoretic arguments~\cite{Calabrese-Gambassi}, 
is that the long wavelength Fourier components dominate the long-time
behaviour of both quantities.  In contrast, 
observables that are sensitive
to the domain wall motion have $X_d^\infty = 0$~\cite{Mayer-etal}, the 
difference residing on the interplay between criticality
and coarsening, a peculiar feature of models with $T_c=
0$~\cite{Corberi1,Mayer-etal}.

The dependence on the initial condition is also interesting.  A
non-zero initial magnetization does not change the value of
$X_s^\infty$ at $T=0$.  Instead, demagnetized initial conditions
 with strong correlations between spins so that only a finite
number of domain walls exist in the system,  yield $X_s^\infty =
0$ (the same result is found in the spherical ferromagnet)~\cite{Sollich1}.
The deviations from non-linear FDTs have not been fully analyzed yet.

The static-dynamics connection~\cite{Frmepape} sketched in
Sect.~\ref{subsub:equilibrium} does not hold in the $1d$ Ising
chain~\cite{Corberi1} and the non-trivial $\chi(C)$ cannot be used to
infer the properties of the equilibrium state. Indeed, the aging part
of the response is finite asymptotically while the equilibrium $P(q)$
has a double-delta (RS) structure as in higher dimensions. The reason
for the failure is that the hypotheses used to 
derive the connection are  not fulfilled.

The large $N$ $O(N)$ model in $D=2$ shares many common features with the 
phenomenology described above~\cite{Chcuyo} although it has not been 
studied in as much detail.

To sum up, a quench to $T=0$ at the lower critical dimension does not
seem to be the dimensional continuation of a line of critical quenches
in the $(T, d) $ plane (as often implicitly assumed), but the
continuation of a line of $T=0$ quenches: the system behaves as in the
coarsening regime, although $X^\infty \neq 0$ for observables that do
not focus on the domain wall dynamics~\cite{Corberi1}.

\subsection{Relaxation in structural glasses}

In particle glassy systems a separation of time-scales exists although
it is not as sharp as in mean-field models or coarsening systems, at
least within simulational and experimental time-scales. In atomic
glasses the existence of an FDT part implies that the rapid particle
vibrations within the cages occur in equilibrium while the structural
relaxation is of a different out of equilibrium kind, and it is not
necessarily ruled by the temperature of the bath. Tests of the
thermodynamic origin of fluctuation-dissipation violations in the
aging regime of these systems were carried through in much more detail
and we summarize them below.

\subsubsection{Simulations of microscopic models}

Mono-atomic and binary Lennard-Jones mixtures, soft sphere systems,
and the BKS potential for silica are standard models for glass forming
liquids.  Both Monte Carlo and molecular dynamics
simulations~\cite{Kob-Barrat,DiLeonardo} suggest that the three first
cases belong to the RFOT class of systems defined in
Sect.~\ref{subsub:parametric} with $T_{\sc eff}=T^{(1)}$ constant in
the aging regime.  $T^{(1)}$ depends weakly on the bath temperature
and systems' parameters but it does not on the preparation protocol as
demonstrated by measurements after quenches and
crunches~\cite{DiLeonardo} or the microscopic
dynamics~\cite{Grigera03}.  Tests of partial equilibration between
fluctuations at different wave-vector gave positive
results~\cite{Kob-Barrat}. Importantly enough, these models have a
well defined equilibrium behaviour and their energy density is
naturally bounded. Of special interest is the numerical method devised
to compute linear responses in molecular systems with high precision
that allowed one to resolve the paradoxical behavior 
previously reported for silica~\cite{Berthier}.

Numerical evidence for a slow decrease in time of the configurational
temperature, as defined in eq.~(\ref{eq:Teff-micro}), although with the
inherent structure complexity, is in
agreement with the idea of the system's representative point
penetrating below the threshold in the (free)-energy landscape~\cite{Kob}.
 
The ratchet effect of an asymmetric intruder in an aging glass was studied 
numerically in~\cite{Gradenigo}. The energy flowing from slow to 
fast modes is rectified to produce 
directed motion. The (sub) velocity of the intruder grows monotonically
with $T_{\sc eff}/T$ and this current could be used to measure $T_{\sc eff}$. 

\subsubsection{Kinetically constrained models}

Kinetically constrained models are toy models of the glassy
phenomenon~\cite{kinetically-constrained,Leonard}.  Their equilibrium
measure is just the Boltzmann factor of independent variables and
correlations only reflect the hard core constraint.  Still, many
dynamic properties of glass forming liquids and glasses are captured
by these models, due to the sluggishness introduced by the constrained
dynamic rules.  The literature on kinetically constrained models is
vast; a recent review with tests of $T_{\sc eff}$
is~\cite{Leonard}. In short, non-monotonic low-temperature response
functions were initially taken as evidence against the existence of
effective temperatures in these systems. The confusion arosed from the
incorrect construction of the $\chi(C)$ plot by using $t_w$ instead of
$t$ fixed (see Sect.~\ref{subsub:pre-asymptotic}) that led to the
incorrect treatment of the transient regime. Still, even this taken
into account, a large number of observables have negative
fluctuation-dissipation ratios; this might be related to the fact that
these models do not have a proper thermodynamics.

\subsubsection{Experiments}

Grigera and Israeloff were the first to measure FDT violations in
glasses by comparing dielectric susceptibility and polarization noise
in glycerol at $T=179.8$K, {\it i.e.}  relatively close to $T_g\simeq
196$K~\cite{Nathan1}. At fixed measuring frequency $\omega\simeq 8$Hz,
they found an effective temperature that slowly diminishes from
$T_{\sc eff} \simeq 185$K to roughly $180$K in $10^5$ sec, that is to
say in the order of days! This pioneering experiment in such a traditional 
glass former has not had a sequel yet.

Particle tracking experiments in a colloidal suspension of PMMA
particles revealed an effective temperature of the order of double the
ambient one from the mobility-diffusivity relation~\cite{Hernan2}.

In the soft matter realm a favorite is an aqueous suspension of
clay, Laponite RG, in its colloidal glass phase. During aging, because
of electrostatic attraction and repulsion, Laponite particles form a
house-of-cards-like structure.  After a number of rather confusing
reports the status of $T_{\sc eff}$ in this system can be summarized
as follows.  The surprisingly high $T_{\sc eff}$ found with dielectric
spectroscopy combined with spontaneous polarization noise measurements
was later ascribed to violent and intermittent events possibly linked
to the presence of ions in the solution which may be the actual source
of FDT violation.  For the moment dielectric degrees of freedom are
invalidated as a good test ground for $T_{\sc eff}$ in this
sample~\cite{Ciliberto}.  Using other methods several groups found
that $T_{\sc eff}$ detaches from the bath temperature.  Strachan {\it
  et al.}~\cite{Strachan} measured the diffusion of immersed probe
particles of different sizes via dynamic light scattering and
simultaneous rheological experiments and found a slightly higher
$T_{\sc eff}$ than $T$.  With micro-rheology Abou and Gallet observed
that $T_{\sc eff}$ increases in time from $T$ to a maximum and then
decreases back to $T$~\cite{Abou}.  Using a passive micro-rheology
technique and extracting $T_{\sc eff}$ from the energy of the probe
particle {\it via} equipartition Greinert {\it et al.} also observed
that $T_{\sc eff}$ increases in time~\cite{Greinert}.  In parallel, a
series of global mechanical tests, and passive and active
micro-rheological measurements that monitor the displacement and
mobility of probe Brownian particles were performed by Ciliberto's and
D. Bonn's groups, both finding no violation of FDT over a relatively
wide frequency range~\cite{Ciliberto,Jabbari,Jop}.  In a very detailed
article Jop {\it et al.} explain many subtleties in the experimental
techniques employed and, especially, the data analysis used to extract
$T_{\sc eff}$ that could have biased the results quoted above.  A
plausible reason for the lack of out of equilibrium signal in some
experiments using Laponite as well as other colloidal
glasses is that the range of
frequency-time explored may not enter the aging regime. Moreover, none
of these works studied the degrees of freedom of the Laponite disks
themselves but, instead, the properties of the solvent molecules or probe
particles. More recently, Maggi {\it et al.}  combined dynamic light
scattering measurements of the correlation function of the colloid
rotations with those of the refringence response~\cite{Maggi} and a
$\chi(C,t_w)$ plot that is rather constant as a function of $C$ and
slowly recovers the equilibrium form as  the
arrested phase is approached ($t_w$ ranges from 90 to 1200 min and the
violations are observed for time differences between 0.1 and 1 ms,
{\it i.e.} frequencies between 10 and 1 kHz). $T_{\sc eff}$ is
at most a factor of $5$ larger than $T$. The actual behaviour of
Laponite remains mysterious -- and not only in what $T_{\sc eff}$ is
concerned!

Oukris and Israeloff measured local dielectric
response and polarization noise in polyvinyl-acetate with
electric-force-microscopy~\cite{Nathan2}. They probed long-lived
nano-scale fluctuations just below $T_g$, achieved a good
signal-to-noise ratio down to very low frequencies, constructed a
parametric plot by keeping $t_w$ fixed and found a non-trivial
asymptotic form with no $t_w$ dependence within the available
accuracy. The data combine into the parametric plot $T_{\sc
  eff}(C) \simeq T C^{-0.57}$ in the aging regime. 

\subsection{Relaxation in frustrated magnetic systems}

Disordered and frustrated magnets behave collectively 
at low temperatures and developed ordered phases that 
although not fully understood are accepted to exist. 
As macroscopic glassy systems they present a separation 
of time-scales in their low-temperature dynamics and 
are good candidates to admit a thermodynamic 
interpretation of the FDT violations.

\subsubsection{Remarks on model systems}

The physics of spin-glasses is a controversial subject.  Some authors
push an Ising domain-growth interpretation of their dynamics -- slowed
down by domain wall pinning by disorder -- a.k.a.  the {\it droplet
  picture}~\cite{Fisher-Huse}. If the scheme discussed in
Sect.~\ref{subsec:coarsening} were reproduced under strong
disorder, the asymptotic $\chi(C)$ plot would have a linear piece of
slope $-1/T$ and a sharp transition at $q_{\sc ea}$ to a flat aging
piece. The domain-growth interpretation is not accepted by other
authors and more complex {\it scenarii} based on the
static~\cite{Mepavi} and dynamic~\cite{Cuku2} solution to the SK model
are envisaged, with a non-trivial $\chi(C)$ as a result. Much effort
has been put in trying to interpret numerical and experimental data as
validating one description at the expense of the other.
Unfortunately, it is very difficult to distinguish between the two. A
third possibility is that, in a loose sense, the spin-glass be like
the low-$T$ phase in the $2d$ XY model, with quasi long-range order.
Yet another proposal is that actual spin-glass samples are of
Heisenberg-type and that chilarity might be decoupled from spin with a
chiral-glass order arriving at a higher critical temperature than
the spin-glass ordering~\cite{Kawamura-model}.

The trap model~\cite{trap} was devised to describe slow dynamics in
systems with weak ergodicity breaking and it was applied, notably, to
describe experiments in spin-glasses. The model shows a glass transition at
a $T_g$ below which an equilibrium Boltzmann state cannot exist. The
$\chi(C)$ has a slope that varies continuously even though there is a
single scaling of relaxation times with age, it depends
non-trivially on the observable and one cannot use it to define a
meaningful $T_{\sc eff}$~\cite{Sollich1}. The reason for this failure
seems to be the unbounded nature of the energy and the
fact that an equilibrium distribution does not exist below
$T_g$.

\subsubsection{Simulations}

Monte Carlo simulations of the $3d$ Edwards-Anderson (EA) model were
carried out by several groups.  One of the hallmarks of the dynamics
of the SK model, dynamic ultrametricity~\cite{Cuku2}, is
absent from all numerical and experimental data analyzed so
far. Magnetic correlation and susceptibility relax in two scales, the
by now usual stationary one for finite time-differences and an aging
one in which the data are well described by a simple $t/t_w$ scaling.
This aging scaling does not conform with the droplet picture either,
which predicts an asymptotic $\ln t/\ln t_w$ form. In all studies
so far, the parametric plot was constructed by keeping $t_w$ fixed and
the curves drift towards increasing values of $\chi$ for longer $t_w$s
as in a transient or critical system.  In simple coarsening problems
the drift with increasing $t_w$ goes in the opposite direction of
rendering the aging part of the curves flatter; this remark suggests
to discard a simple droplet picture. The
outcome $\chi(C)$ found for the longest $t_w$ reached was interpreted
as being non-constant~\cite{Juan} -- as in the SK model -- although
this is, in our opinion, not that clear from the data that could
be described by a straight line.  The simultaneous
$t/t_w$ scaling, the lack of unambiguous evidence for a stable plateau at 
$q_{\sc ea}$, and a curved $\chi(C)$ in the
aging regime is not what would be expected from an analogy with the SK
model. Instead, it would be consistent with critical dynamics and the $2d$ XY model
similitude.  A number of caveats on the numerical analysis should,
however, be lifted before reaching a firm conclusion. 

The finite-time finite-length relation between static
$\aleph(C,\xi(t_w))$ and long-time out of equilibrium dynamic
$\chi(C,t_w)$ (Sect.~\ref{subsub:equilibrium})~\cite{Barrat-Berthier}
was put to the test in the $2d$ and $3d$ EA models at finite $T$.  The
notable coincidence of the two functions found in the $2d$ case, in
which there is no complex equilibrium structure, suggests that the
claimed coincidence of $\chi(C)$ and $\aleph(C)$ in $3d$~\cite{Juan}
might also be valid {\it just} in the transient regime.

Simulations of the $3d$ Heisenberg spin-glass model with weak
anisotropy suggest that $T_{\sc eff}$ associated to the spin degrees
of freedom is constant and about twice the critical temperature for
spin-glass ordering~\cite{Kawamura-FDT}. As far as we know, chiral degrees
of freedom have not been used to estimate $T_{\sc eff}$. 

As regards fluctuations, the two kinds discussed in
Sect.~\ref{sec:local-measurements} were measured in the $3d$ EA
spin-glass.  Disordered induced ones~\cite{Roma}, in which one
computes strictly local noise-averaged correlations and linear
responses, demonstrate the existence of two types of spins in each
sample: rapid paramagnetic-like ones and slow ones. The former satisfy
FDT while the latter evolve in two time-regimes with a fast one
satisfying FDT and a slow one in which $\chi_i(C_i)$ looks quite flat
as in coarsening systems. The simulation suggests that 
the two ensembles behave independently of
each other and are strongly correlated with the backbone of the ground
state configurations.  The average over all sites (at finite $t_w$)
gives rise to a curve with non-constant slope. These results suggest a
still different picture for the spin-glass dynamics
in which a rather compact set of spins undergoes 
coarsening of the backbone equilibrium configurations while the other 
ones behave paramagnetically. This intriguing idea
needs to be put to further test.

The analysis of noise induced fluctuations suggests that 
eq.~(\ref{eq:local-T}) is valid although better numerical
data would be needed to have definitive evidence for
this statement. A more detailed discussion can be found 
in~\cite{Chamon-Cugliandolo}. Very recent studies of non-linear fluctuations 
that take advantage of FDRs to compute higher order responses point
in the direction of the TRI scenario~\cite{Corberi11} with a 
finite $T_{\sc eff}$. 

\subsubsection{Experiments}

On the experimental side the first attempt to quantify FDT violations
in spin-glasses was indirect~\cite{Grempel}.  Simultaneous
measurements of global magnetic noise and susceptibility in the 
thiospinel insulating spin-glass  were later performed
by H\'erisson and Ocio~\cite{Ocio}.  The data confirm deviations from
the FDT with a $\chi(C,t_w)$ plot of relatively curved form although
still evolving during the experimental time window.  The authors
interpreted it as evidence for the full RSB scenario, {\it via} the
association $\chi(C) \leftrightarrow \aleph(C)$. However, as with
numerical data~\cite{Juan}, dynamic ultrametricity fails to show off,
the asymptotic limit of the parametric construction is still far, and
a clear-cut distinction between a curved and a linear $\chi(C)$ is
hard to assess. 

More recent experiments exploit two novel techniques,
Hall-sensor based magnetometer and giant magnetoresistance technology
to detect signals from very small samples~\cite{Denis-Lhote}. The use
of these probes opens the way to perform a systematic study of FDT
violations in magnetic systems of different kind (spin-glasses,
super-spin glasses, disordered ferromagnets...).  The first of these
measurements appeared recently~\cite{Denis-Lhote2} in a super-spin
glass, a system of magnetic nanoparticles suspended in fluid
glycerol with a single-domain magnetic structure that behaves as one
large spin, the orientation of which is the only degree of freedom.
The large magnetic moment facilitates the observation of magnetic
noise.  For aging times of the order of 1 h, the ratio of $ T_{\sc
  eff}$ to the bath temperature T grows from 1 to 6.5 when T is
lowered from $T_g$ to $0.3 \ T_g$, regardless of the noise frequency.

Artificial spin ice is yet another material in which the $T_{\sc eff}$ 
notion has been tested~\cite{spin-ice}.

\subsection{Driven liquids and glasses}
\label{subsec:driven-numerical}

In~\cite{Berthier-Barrat} the molecular dynamics of a binary
Lennard-Jones mixture under a steady and homogeneous shear flow was
studied. The deviation from FDT is similar to the one found
analytically in disordered spin models of RFOT type with asymmetric
couplings that mimic non-conservative
forces~\cite{Cukulepe,Bebaku}. Moreover, it does not depend on the
observable.  The tracer particle experiment was also realized.  When
the tracers' Einstein frequency is smaller than the inverse relaxation
time of the fluid, a non-equilibrium equipartition theorem holds with
$m_{\sc tr}v^2_z= T_{\sc eff}$, where $v_z$ is the velocity in the
direction transverse to the flow. For increasing $m_{\sc tr}$ the
effective temperature very slowly crosses over from $T$ to the slow
modes value, in perfect agreement with the notion of a
temperature measured by a thermometer sensible to the scale. $T_{\sc
  eff}$ also captures the essential phenomenological idea that when a
system is sheared more vigorously its effective temperature increases.

O'Hern {\it et al.} also studied fluctuation-dissipation relations in
shear fluids~\cite{OHern}. This group defined an effective temperature
through the `static limit' $\lim_{t\to\infty}\chi(t-t_w)/C(t,t)$, a
kind of average of the slope of the $\chi(C)$ plot over the full range
of $C(t-t_w)$ that mixes different time scales (in particular, the
high and low frequency ones).  A more thorough
discussion of the comparison between this definition and the one
described in this review was given by Ilg and Barrat~\cite{Ilg2}
within a fully solvable model that demonstrates the
importance of {\it not} mixing time-scales to get physically sensible
results.

A first study of the fluctuations of entropy production in a Lennard-Jones
fluid above and below $T_g$ under a shear flow appeared in~\cite{Angelani05}
and the need to take into account $T_{\sc eff}$, as obtained from 
the modification of the FDT below $T_g$, was signaled in this 
paper. A more detailed analysis of the time-scale dependent
effective temperature would be needed to fully test the proposal 
in~\cite{Bonetto}.

Another prominent example is the current driven motion of vortices in
type II superconductors. Disorder reduces dissipation, is responsible
for non-equilibrium transport and magnetic properties. The external
force induces two dynamic phase transitions separating plastic flow,
smectic flow and a frozen transverse solid.  A low-frequencies $T_{\sc eff}$ that
decreases with increasing driving force and reaches the equilibrium
melting temperature when the dynamic transverse freezing occurs was
computed from the transverse motion in the fluid moving
phase~\cite{Kolton1}.

\subsection{Granular matter}

Several studies of the effective temperature of granular matter have
been pursued theoretically, numerically and experimentally. In the
latter front, D'Anna {\it et al.} immersed a torsion oscillator in a
granular system fluidized by strong high frequency external vibrations
to realize the `thermometer' experiment. They found $T_{\sc eff}
\propto \Gamma^2$ with $\Gamma$ the adimensional measure of
vibrational intensity, and quite independently of $\omega$~\cite{Danna}.  Wang {\it
et al.}~\cite{Wang} visualized the dynamics of tracer particles
embedded in a $3d$ granular ensemble slowly sheared by the rotating
inner wall of a Couette cell. $T_{\sc eff}$, as obtained from the
comparison between the tracer's diffusion and mobility perpendicular
to the applied rate of strain, is independent of the shear rate used and
the tracers properties but does depend on the packing density of the
system. Tests of the thermodynamic properties of $T_{\sc eff}$ have
not been carried through in this system yet.  The dependence on the
direction of the applied stress was studied by Twardos and Dennin in a
plastic bead raft close to jamming~\cite{Dennin}.  As expected, the
correlations and linear responses in the direction of flow do not
decay slowly and $\chi(C)$ does not have the same properties as in the
transverse direction (cfr.~\cite{Kolton} and~\cite{Jorge-Hernan}).
Gei and Behringer stressed the fact that in a granular assembly the
outcome of a mobility measurement depends on whether one imposes the
velocity or the external force~\cite{Behringer}.

In the powders literature reference is often made to the `granular
temperature', a measure of the temperature of the fast modes, as given
by the kinetic energy of the grains
$T_K\equiv \frac{2}{d} E_K \equiv \frac{2}{d} \langle v^2\rangle$.
Importantly enough, $T_K$ is a high frequency measure that does not
really access the structural properties of the sample and, in a sense,
plays the role of the environmental temperature in thermal systems.
$T_K$ is generically smaller than $T_{\sc eff}$, as in thermal systems
where $T_K=T$, the temperature of the bath.

\subsection{Activated dynamics}

Activated processes often occur in systems that are out of
equilibrium, in the sense that their response to an external drive is
strongly non-linear or that their phase space distribution is not the
Gibbs-Boltzmann one. The question as to whether an Arrhenius law
governs the activation rate, possibly with an effective temperature,
and how the latter compares to the one defined from the deviations
from FDT has been addressed recently~\cite{Ilg1}. Ilg and Barrat
studied the effect of an out of equilibrium flowing environment, a
weakly sheared super-cooled liquid, on the activated dynamics 
between the two stable conformations of dumbbell particles. The
transition rate is well described by an Arrhenius law with a
temperature that crosses over from the one of the equilibrium bath to
a higher value close to the $T_{\sc eff}$ of the slow modes of the
driven fluid. The crossover roughly occurs at the value of the rate
that corresponds to the inverse of the $\alpha$ relaxation time of the
fluid.

Three related studies are also worth mentioning. An effective
temperature, also consistent with the one stemming from
fluctuation-dissipation measurements, appears in a phenomenological
Arrhenius law that describes transverse jumps between channels in the
driven motion of vortex lattices with random pinning~\cite{Kolton}.
Haxton and Liu showed that in the shear dominated regime the stress of
a $2d$ sheared fluid follows an Arrhenius law with the effective
temperature~\cite{Haxton}. A study of activation and $T_{\sc
  eff}$ in a $2d$ granular system close to jamming appeared
in~\cite{Abate}.

\subsection{Biological systems}

In biologically inspired problems the relevance of $T_{\sc eff}$ was
stressed to reveal the active process in hair bundles~\cite{Martin}
and model cells~\cite{mizuno07}.
Morozov {\it  et al.}~\cite{Morozov} studied a model of the cytoskeletal network
made of semi-flexible polymers subject to thermal and motor-induced
fluctuations and found a $T_{\sc eff}$ that exceeds the environmental
temperature $T$ only in the low-frequency domain where motor agitation
prevails over thermal fluctuations. 
Simple gene network models were
studied from the $T_{\sc eff}$ perspective in~\cite{Lu}.
Fluctuation-dissipation ratios were used to quantify the degree of
frustration, due to the existence of many metastable disordered
states, in the formation of viral capsids and the crystallization of
sticky discs, two self-assembly processes~\cite{Jack}.   Fluctuations and responses of
blood cell membranes for varying ATP concentration were measured very
recently~\cite{Yair}. The measured $T_{\sc eff}$ approaches the bath
temperature at high frequencies and increases at low frequencies
reaching 4-10 times the ambient temperature.

Ratchets are simple models of molecular motors, 
out of equilibrium systems with directed dissipative
transport in the absence of any external bias. 
Harada and Sasa proposed
to use the violations of FDT in flashing ratchets as a means to
measure the energy input per unit time in molecular motors -- an
otherwise difficult quantity to access~\cite{Harada-Sasa}. Kolton
showed that the rectified transverse velocity of a driven particle in a
geometric ratchet is equivalent to the response of a $1d$ flashing
ratchet at a drive-dependent $T_{\sc eff}$, as defined from
the generalized Einstein relation~\cite{Kolton-ratchet}. 

Active matter is driven out of equilibrium by internal or external
energy sources. Its constituents absorb energy from their environment
or from internal fuel tanks and dissipate it by carrying out internal
movements that lead to translational or rotational motion. A 
typical example are self-propelled particle assemblies in bacterial colonies.
The role played by $T_{\sc eff}$ in the stability of dynamic phases of motorized particle
systems was stressed by Shen and Wolynes~\cite{Shen}. Multiple
measurements of $T_{\sc eff}$ were carried out with
molecular dynamic simulations of motorized spherical as well as linear
molecules in interaction~\cite{Loi}. All measurements (from
fluctuation-dissipation ratio and using tracers) yield a constant 
low-frequency
$T_{\sc eff}>T$ when the effect of the motors is not
correlated with the structural rearrangements they induce. Instead, $T_{\sc eff}$
takes a slightly lower value than $T$ when susceptible motors are used, as argued
in~\cite{Shen}. Such an `inversion'  also occurs in relaxational systems in which
the initial configuration is chosen to be 
one of equilibrium at a lower $T$ than the working one~\cite{Behose,Ig,Krzakala}.
In the case of uncorrelated motors, $T_{\sc eff}/T$ was found to follow the empirical law
$T_{\sc eff}/T \simeq 1+\gamma f^2$ with $f$ the active force relative to 
the mean potential force and  $\gamma\sim 15$ a parameter.
Palacci {\it et al.}.~\cite{palacci10} 
investigated $T_{\sc eff}$  by following 
Perrin's analysis of the density profile in the steady state of an active colloidal 
suspension under gravity. The active particles used -- JANUS particles -- are 
 chemically powered colloids 
and the suspension was studied  with optical microscopy. The measurements show that the 
active colloids are hotter than in the passive limit with a
$T_{\sc eff}$ that increases as the square of the parameter that controls 
activation, the Peclet number, a dependence that is highly reminiscent 
of the $f^2$-dependence of the simulations mentioned above.
Other theoretical studies of $T_{\sc eff}$ in active matter appeared in~\cite{Tailleur}.

Joly {\it et al.}~\cite{joly10} used numerical techniques to study the non-equilibrium steady 
state dynamics of a heated crystalline nanoparticle  suspended in a fluid. 
This problem models an active colloid that acts as a local heat source and
generates a temperature gradient around it.  
By comparing the mobility 
to the velocity correlation function, they found that the FDT
approximately holds at short-time lags with a temperature value that 
coincides with the kinetic one. In contrast, at long-time lags data are
compatible with the temperature estimated by using the Einstein relation.

Certainly, many more studies of effective temperatures will appear in this 
very active field of research, essentially out of equilibrium, in the near
future.

\subsection{Plasticity}

Langer~{\it et al.} extended the traditional phenomenological
defect-flow theory to the shear transformation zone (STZ) theory of
large-scale plastic deformation in amorphous materials~\cite{Langer}.
The new theory incorporates effective temperature ideas. It is a picture of
plastic deformation in molecular glasses in which a `disorder
temperature' characterizes the steady state of the system and 
controls the slow processes. The relation
between the disorder temperature and a configurational entropy, under
the assumption of a sharp separation of time scales between structural
and vibrational processes, as well as other
thermodynamic properties of it -- along the lines discussed in
Sect.~\ref{sec:landscape} -- were addressed. As far as we know, there
have been no tests to compare the STZ disorder temperature and the 
effective temperature computed from fluctuation-dissipation measurements;
this is the reason why we keep distinct names for the two quantities.
The STZ theory suggests an explanation of shear-banding instabilities that 
have been put to the numerical test in~\cite{Shi}.

Another approach to plasticity consist in adapting the RFOT-replica 
approach to this problem~\cite{Yoshino10}. Numerical simulations 
of binary soft sphere mixtures at low temperature prove that the stress relaxation and its
response to a strain step are also linked by a modified FDT with a single valued 
$T_{\sc eff}$ in the aging regime~\cite{Yoshino}. 

\subsection{Turbulent fluids}

Last but not least, the experimental quest for effective temperatures
in the steady state of turbulent flows has recently restarted.  As far
as we know, two experiments on turbulent flows appeared in the
literature after the r\^ole played by different time-scales was
stressed in the investigation of effective temperatures in macroscopic
systems with slow dynamics.
 
The transverse fluctuations and the response of a string held at its
ends and at constant tension (a mechanical probe) in the inertial
regime of a stationary turbulent air jet flow were compared to obtain
the effective temperature. The ranges of wave-numbers, and Reynolds
number ($7.4 \ 10^4 \leq Re \leq 1.7 \ 10^5$).  accessed in this
experiment are pretty wide~\cite{Naert}. All measurements are
compatible with $T_{\sc eff} \propto k^{-11/3}$. The $k$ dependence
confirms that there is no equilibrium between Fourier modes due to the
energy flux between scales. The particular exponent is explained
in~\cite{Naert} with a simple model derived from Kolmogorov 1941.

In a more recent  experiment focus was set on an anisotropic
non-homogeneous  axisymmetric von K\'arm\'an flow at large 
Reynolds number~\cite{Daviaud}. The measurements are, though, somehow 
indirect and yield an effective temperature 
that depends  on the observable. This feature 
needs to be better characterized.

\subsection{Quantum models}
\label{subsec:quantum}

There is growing interest in the dynamics of quantum systems. Of
particular importance in this field is to distinguish cases in which
the system of interest is isolated and the dynamics occur at constant
energy from those in which the system is coupled to an environment and
the dynamics are dissipative. The out of equilibrium dynamics are
typically induced in two ways: the system is driven out of equilibrium
by, for instance, a coupling to electric or heat current sources, or
it is strongly perturbed by time-dependent external fields; the system
evolves after a quench meaning that a parameter in its Hamiltonian is
changed with some protocol.  All these cases are easily realized in
the laboratory nowadays and their potential applicability is being
explored.  Theoretical searches for effective temperatures in these
problems, defined in different ways, are starting to appear in the
literature.

As recalled in Sect.~\ref{sec:definitions} linear responses and
correlation functions in equilibrium quantum systems are related by
FDTs that take slightly different form depending on the bosonic or
fermionic character of the observables. In an out of
equilibrium situation one can compare the linear response and the
correlation and test whether, at least in some dynamic regime, a
parameter replacing the environmental temperature appears. This route
was first taken within mean-field quantum glassy models quenched 
from their disordered into
their ordering phase~\cite{Culo,Malcolm,Arbicu2}. In these models, the
quantum FDT holds within the rapid stationary scale.  Instead, when
the dynamics gets slow and the aging regime is attained the relation
between linear response and correlation takes
the classical form, with an effective temperature that  is
higher than the one of the bath and different from zero. As in the 
classical case, the static properties of these models can be solved
with the replica and TAP approaches and a connection with the 
dynamic solution can be established~\cite{static-quantum}.

More recently, the analysis of mesoscopic quantum models commenced.  A
metallic ring threaded by a time-dependent magnetic field and coupled
to a lead in equilibrium at temperature $T$ and chemical potential
$\mu$ was studied in~\cite{Lili1}. The numerical solution to the
Schwinger-Keldysh equations for the free but driven fermions, in the
limit of small dissipation, suggests that their Green functions
satisfy an FDT with constant $T_{\sc eff}$. More recently, Arrachea
{\it et al.}  attacked the problem of a wire connected to left and
right reservoirs (in equilibrium at the same chemical potential and
temperature) and driven out of equilibrium by different ac pumps
locally connected to the wire~\cite{Lili2}. The local effective
temperature was computed from the modified FDT and by requiring that
there be no heat flow to nor from  macroscopic probes, {\it i.e.} thermometers,
weakly coupled to chosen sites on the device.  For weak driving and
environmental temperatures lower than the Fermi energy of the
electrons these two measurements coincide on each site.  $T_{\sc eff}$
has spatial $2k_F$-Friedel-like oscillations (allowing for local cooling) but its spatial 
average
is higher than the temperature of the reservoirs when the pump is
applied. Moreover, the direction of heat flow between the device and 
each one of the leads is dictated by the 
temperature gradient at the contact, defined as the 
difference between $T_{\sc eff}$ at the contact and the temperature
of the lead. For conveniently chosen pumps the device can extract 
energy from one lead and transport it to the opposite one.

The theory of isolated quantum models is a very
active field of research, pushed by the large number of experiments
on cold-atom systems that are being realized World-wide. Questions 
on thermalization after a quantum quench, in which a parameter in 
the Hamiltonian is suddenly changed, are being posed, extending this
very much investigated issue in classical statistical mechanics to quantum 
systems. At present the belief is that non-integrable quantum systems
do reach equilibrium while integrable ones do not. Rigol {\it et al.}
proposed to use extended Gibbs-Boltzmann measures in the latter case
with as many constants of motion as necessary fixed by the initial 
condition~\cite{Rigol}.
Currently, this proposal is been debated and studied in particular cases.
For instance, Rossini {\it et al.} complemented it with the idea that 
non-local  observables (when expressed in terms of quasi-particles)
should equilibrate in integrable systems while local ones should not 
under general 
conditions~\cite{Rossini}. They  based this conjecture on their mixed analytic/numeric 
solution of a quantum spin chain. Studies of effective temperatures (as defined
from deviations from FDT) in this context
are under way~\cite{Foini}. Another quantum setting in which an effective temperature
was proposed is the entanglement entropy of a 
sub-system~\cite{Racz}.

\section{Conclusions}

In this review we discussed the notion of an effective
temperature~\cite{Hosh}-\cite{Cuku3} that stemmed from the deviations from the
fluctuation-dissipation theorem found in the analytic
solutions to simple glass models~\cite{Cuku1,Cuku2} and
self-consistent approximations to more realistic
glassy systems~\cite{PhysicaA,Cavagna}.  Although phenomenological
definitions of out of equilibrium temperatures were not lacking in the
literature, see {\it e.g.}~\cite{To,Jou}, the
effective temperature discussed in this report lies, in our opinion, on 
a firmer physical basis.

Fluctuation dissipation violations can be searched for in all out of
equilibrium system. Whether the outcome can be interpreted as giving
origin to an effective temperature is a different and more delicate issue.
In this review we tried to distinguish cases in which the latter can
be done from cases in which it is not possible. A necessary condition
seems to be that the system should evolve slowly, irrespectively of
whether it relaxes or is in a non-equilibrium steady state, in a small entropy
production regime~\cite{Cuku3}. Although not rigorously proven 
it seems natural to require that it approaches an 
approximately flat measure on a region of phase space, the entropy of 
which (also called  complexity)
should give an alternative, microcanonic-like, access to the effective
temperature~\cite{Edwards,Remi,Martens,Nieuwenhuizen}.  
A consequence of the last condition is that
the energy density, and more generally the
averages of one-time dependent observables, converge to {\it finite}
values. All these features are
satisfied exactly in some solvable mean-field-like glasses such as the
celebrated $p$-spin disordered model and the low-$T$ mode-coupling
approximations.  Numerical simulations suggest they also are, at least
within a given time regime, in a large number of glassy systems with
short-range interactions, including Lennard-Jones mixtures and
others. In all these cases the systems have a rather complex
collective behavior with a separation of time-scales fast-slow that
has an influence in the values that the effective temperature takes.

As far as it has been checked, in all the above-mentioned cases the
effective temperature conforms to the common prejudices one has of a
temperature and, more importantly, it has the most welcome property of
being measurable directly, hence being open to straightforward --
though difficult to implement -- experimental tests.  Central to the
correct identification is the realization that the relaxation
time-scales have to be correctly identified and that measurements have
to focus on each of them separately.
Simulations in atomic glasses are very complete and yield pretty
convincing support to the effective temperature
ideas~\cite{Kob-Barrat}-\cite{Gradenigo},\cite{Berthier-Barrat}-\cite{Angelani05}.
 Experimental subtleties
prove to be difficult to keep under control, especially in colloidal
suspensions~\cite{Hernan2}-\cite{Maggi}, 
but successful measurements of fluctuation-dissipation violations in structural
glasses~\cite{Nathan1,Nathan2} have been surveyed.
Very recent numerical~\cite{Loi} and experimental~\cite{palacci10}  studies 
of active matter are also very promising.
Simulations of spin-glass~\cite{Juan} clearly demonstrate the violation of 
FDT but are not as precise as to determine beyond doubt  
the actual functional form of the modified relation. In
our opinion, pre-asymptotic effects are still to be disentangled from
the asymptotic regime in a satisfactory way. The same applies to 
the first experiments in an insulator spin-glass~\cite{Ocio} the outcome of which could be 
greatly improved by the use of innovative 
techniques~\cite{Denis-Lhote,Denis-Lhote2}. A careful analysis of the 
thermodynamic properties of the fluctuation-dissipation ratio remains 
to be done such frustrated and disordered magnets.

Quite naturally, the idea that effective temperatures could be relevant to
other out of equilibrium systems was explored in the last fifteen years. 
Critical quenches in dissipative classical systems~\cite{Calabrese-Gambassi}, zero temperature 
dynamics at the lower critical dimension~\cite{Corberi1}, 
granular matter~\cite{Jorge-Edwards}-\cite{Bertin-Dauchot} and turbulent 
fluids~\cite{Naert,Daviaud} are just some examples  in which fluctuation dissipation relations 
have been studied. The question remains as to whether 
the outcome admits a thermodynamic interpretation in, at least, some 
limits. In the case of critical quenches, this seems to be the case when the 
long two-time limit prescription of Godr\`eche and Luck is 
taken~\cite{Godreche-Luck-critical}. Simulations of 
weakly perturbed granular matter are very encouraging~\cite{Jorge-Hernan}
but experiments have turned out to be much harder to realize~\cite{Dauchot}
(see, though,~\cite{Wang}).
In turbulent fluids the question remains open experimentally.

Out of equilibrium quantum systems are receiving enormous attention 
nowadays, boosted by experiments in cold atoms and nanotechnology. 
In this realm, fundamental questions of thermalization arise,
especially in isolated samples submitted to a quench. The effective temperature 
has been studied in a few dissipative quantum systems out of equilibrium,
both mean-field~\cite{Culo,Malcolm} and low-dimensional~\cite{Lili1,Lili2}.
 The analysis of whether it also plays a r\^ole in quantum quenches of isolated
 samples is under way~\cite{Foini}.

This article presents a vast panorama of fluctuation dissipation deviations 
in non-equilibrium classical, and to a much smaller extent,
quantum systems, and their interpretation, in some cases, in terms
of effective temperatures. Many questions of fundamental interest remain 
open in this field. One of the main open challenges in the 
context of glassy systems, is to find the  microscopic 
origin of these modifications. From a wider viewpoint the validity of the 
effective temperature concept should be more clearly delimited. 

\vspace{0.5cm}

\noindent{\it Acknowledgments.} I especially wish to thank J. Kurchan and
L. Peliti for our collaboration on the definition and study of the
effective temperatures. I also wish to thank J. Langer for inducing me  
to write this overview. This work has been supported by ANR-BLAN-0346.

\vspace{0.5cm}


\begin{thebibliography}{99}

\bibitem{Hosh} P. C. Hohenberg and B. I. Shraiman, 
{\it Chaotic behavior of an extended system}, 
Physica D {\bf 37}, 109  (1989). 

\bibitem{Edwards} S. F. Edwards, {\it The role of entropy in the specification of
a powder}, in Granular matter, an interdisciplinary approach, A. Mehta ed. 
(Spinger-Verlag, New York, 1994). S. F. Edwards and D. V. Grinev, 
{\it Statistical mechanics of stress transmission in disordered granular arrays}, Phys. Rev. Lett. 
{\bf 82}, 5397 (1999). 

\bibitem{Cukupe} L. F. Cugliandolo, J. Kurchan, and L. Peliti,
{\it Energy flow, partial equilibration, and effective temperatures in systems with slow dynamics},
 Phys. Rev. E {\bf 55}, 3898 (1997).

\bibitem{Cuku3}
L. F. Cugliandolo and J. Kurchan,
{\it A scenario for the dynamics in the small entropy production limit},
 J. Phys. Soc. Japan {\bf 69},  247 (2000).

\bibitem{Cuku1}
L. F. Cugliandolo and J. Kurchan, {\it Analytical solution of the off-equilibrium dynamics of a
 long-range spin-glass model}, Phys. Rev. Lett. {\bf 71}, 173 (1993).

\bibitem{Cuku2}
L. F. Cugliandolo and J. Kurchan, {\it The out of equilibrium dynamics of the
Sherrington-Kirkpatrick model}, J. Phys. A {\bf 27}, 5749 (1994). 

\bibitem{Remi}
R. Monasson,
{\it  Structural Glass Transition and the Entropy of the Metastable States}, 
 Phys. Rev. Lett. {\bf 75}, 2847 (1995).

\bibitem{To} A. Q.  Tool, 
{\it Relation between inelastic deformability 
and thermal expansion of glass in its annealing range}, 
J. Am. Ceram. Soc. {\bf 29}, 240 (1946).

\bibitem{fictrev}
J. J\"ackle, {\it Models of the glass-transition},
Rep. Prog. Phys. {\bf 49}, 171 (1986).
G. W. Scherer, {\it  Theories of relaxation},  J. Non Cryst. Solids {\bf 123}, 75  (1990). 
I. M. Hodge,  {\it Enthalpy relaxation and recovery in amorphous materials},
J. Non Cryst. Solids {\bf 169}, 211 (1994).

\bibitem{Jou}
D. Jou, J. Casas-V\'azquez and G. Lebon, 
{\it Extended Irreversible Thermodynamics} (Springer, Berlin, 1993). 

\bibitem{Crri} A. Crisanti and F. Ritort, {\it Violation of the
    fluctuation-dissipation theorem in glassy systems: basic notions
    and the numerical evidence}, J. Phys. A {\bf 36}, R181
  (2003).
  
  \bibitem{Calabrese-Gambassi}
P. Calabrese and A. Gambassi,
{\it Ageing Properties of Critical Systems},
 J. Phys. A {\bf 38}, R133 (2005).

  \bibitem{Corberi1}
F. Corberi, E. Lippiello, and M. Zannetti,
{\it Fluctuation-Dissipation relations far from Equilibrium},
 J. Stat. Mech. P07002 (2007). 

\bibitem{Leuzzi}
L. Leuzzi,
{\it A stroll among effective temperatures in aging systems: limits and perspectives}, 
J. Non-Cryst. Solids {\bf 355}, 686 (2009). 


\bibitem{Semerjian} 
G. Semerjian, L. F. Cugliandolo, and A. Montanari, 
{\it On the stochastic dynamics of disordered spin models}, 
J. Stat. Phys. {\bf 115}, 493 (2004).

\bibitem{Vulpiani}
U. Marini Bettolo Marconi, A. Puglisi, L. Rondoni, and A. Vulpiani,
{\it Fluctuation-Dissipation: Response Theory in Statistical Physics},
Phys. Rep. {\bf 461}, 111 (2008). 

\bibitem{Cukupa}
L. Cugliandolo, J. Kurchan, and G. Parisi, 
{\it Off equilibrium dynamics and aging in unfrustrated systems}, 
J. Phys. I France {\bf 4}, 1641 (1994).

\bibitem{Arbicu} C. Aron, G. Biroli, and L. F. Cugliandolo, 
{\it Symmetries of generating functionals of Langevin processes 
with colored multiplicative noise}, J. Stat. Mech. P11018 (2010).

\bibitem{Speck}
M. Baiesi, C. Maes, and B. Wynants,
{\it Nonequilibrium linear response for Markov dynamics, I: jump 
processes and overdamped diffusions}
J. Stat. Phys. {\bf 137}, 1094 (2009).
M. Baiesi, E. Boksenbojm, C. Maes, and B. Wynants,
{\it Nonequilibrium linear response for Markov dynamics, II: inertial dynamics},
J. Stat. Phys. {\bf 139} 492  (2010).
T. Speck, {\it Driven soft matter: entropy production and the fluctuation-dissipation theorem},
Prog. Theor. Phys. Suppl. {\bf 184}, 248 (2010).

\bibitem{Prost}
J. Prost, J-F Joanny, and J. M. R. Parrondo,
{\it Generalized fluctuation-dissipation theorem for steady-state
systems}, Phys. Rev. Lett. {\bf 103}, 090601 (2009).

\bibitem{Martens}
K. Martens, E. Bertin, and M. Droz,
{\it Dependence of the fluctuation-dissipation 
temperature on the choice of observable},
Phys. Rev. Lett. {\bf103}, 260602 (2009);
{\it Entropy-based characterizations of the observable-dependence of the 
fluctuation-dissipation temperature},
Phys. Rev. E {\bf 81}, 061107 (2010).

\bibitem{SchwingerKeldysh} 
U. Weiss, {\it Quantum dissipative systems} (World Scientific, Singapore, 1993).

\bibitem{Kamenev}
A. Kamenev, arXiv:cond-mat/0412296, arXiv:cond-mat/0109316.

\bibitem{Gawedski} R. Chetrite and K. Gawedzki,
{\it Eulerian and Lagrangian pictures of non-equilibrium diffusions}, 
J. Stat. Phys. {\bf 137}, 890 (2010). 

\bibitem{Lippiello} 
E. Lippiello, F. Corberi, A. Sarracino, and M. Zannetti, 
{\it Nonlinear response and fluctuation dissipation relations}, 
Phys. Rev. E {\bf 78}, 041120 (2008).

\bibitem{book-lengths}
{\it Dynamical heterogeneities, glasses and granular media},
ed. by L. Berthier, G. Biroli, J.-P. Bouchaud, L. Cipeletti and W. van Saarloos
(Oxford Univ. Press, Oxford, 2011). 



\bibitem{KTW1}
T. R. Kirkpatrick and D. Thirumalai, 
{\it p-spin-interaction spin-glass models - connections with the structural glass problem},
Phys. Rev. B {\bf 36}, 5388 (1987). 

\bibitem{KTW2} T. R. Kirkpatrick and P. G. Wolynes, 
{\it Stable and metastable states in mean-field potts and structural glasses},
Phys. Rev. B {\bf 36}, 8552 (1987). 

\bibitem{KTW3}
 T. R. Kirkpatrick, D. Thirumalai, and P. G. Wolynes, 
{\it Scaling concepts for the dynamics of viscous-liquids near an ideal glassy state},
Phys. Rev. A {\bf 40}, 1045 (1989).

\bibitem{Mepavi}
M. M\'ezard, G. Parisi, and M. A. Virasoro,
{\it Spin glass theory and beyond}, 
(World Scientific, Singapore, 1987). 

\bibitem{TAP}
D. J. Thouless, P. W. Anderson, R. G. Palmer, 
{\it Solution of solvable model of a spin glass},
Phil. Mag. {\bf 35}, 593 (1977). 

\bibitem{PhysicaA}
J-P Bouchaud, L. F. Cugliandolo, J. Kurchan, M. M\'ezard
{\it Mode-Coupling approximations, glass theory and disordered systems},
Physica A {\bf 226}, 243 (1996).

\bibitem{Cavagna}
V. Lubchenko and P. G. Wolynes,
{\it Theory of structural glasses and supercooled liquids}
Annual Rev. Phys. Chem. {\bf 58}, 235 (2007). 
A. Cavagna, {\it Supercooled liquids for pedestrians},
Phys. Rep. {\bf 476}, 51 (2009). 
L. Berthier and G. Biroli, {\it  theoretical perspective on the glass transition and nonequilibrium phenomena in disordered materials},  arXiv:1011.2578.


\bibitem{Cukule}
L. F. Cugliandolo, J. Kurchan, and P. Le Doussal, 
{\it Large time out-of-equilibrium dynamics of a manifold in a 
random potential}, 
Phys. Rev. Lett. {\bf 76}, 2390 (1996).

\bibitem{Frme}
S. Franz and M. M\'ezard, 
{\it On mean-field glassy dynamics out of equilibrium}
Physica A {\bf 210}, 48 (1994);
{\it Off-equilibrium glassy dynamics - a simple case}, 
Europhys. Lett. {\bf 26}, 209 (1994). 

\bibitem{Culo}
L. F. Cugliandolo and G. Lozano,
{\it Real-time non-equilibrium dynamics of quantum glassy systems},
Phys. Rev. B {\bf 59}, 915 (1999);
{\it Quantum aging in mean-field models},
 Phys. Rev. Lett. {\bf 80}, 4979 (1998). 

\bibitem{Malcolm}
M. P. Kennett and C. Chamon, 
{\it Time reparametrization group and the long time behaviour in quantum glassy systems}, 
Phys. Rev. Lett. {\bf 86}, 1622 (2001).
M. P. Kennett, C. Chamon, and Y. Yu, 
{\it Aging dynamics of quantum spin glasses of rotors}, 
Phys. Rev. B {\bf 64}, 224408 (2001).

 \bibitem{Sollich1}
  P. Sollich, S. Fielding, P. Mayer,
 {\it Fluctuation-dissipation relations and effective temperatures 
in simple non-mean field systems}, 
J. Phys.: Condens. Matt. {\bf 14}, 1683 (2002).  


\bibitem{Frmepape}
S. Franz, M. M\'ezard, G. Parisi, and L. Peliti, 
{\it The response of glassy systems to random perturbations: A bridge between equilibrium 
and off-equilibrium}, 
J. Stat. Phys. {\bf 97}, 459 (1999);
{\it Measuring equilibrium properties in aging systems}, 
Phys. Rev. Lett. {\bf 81}, 1758 (1998). 

\bibitem{Mepa}
M. M\'ezard and G. Parisi,
{\it Statistical physics of structural glasses}, 
J. Phys.: Condens. Matt. {\bf 12}, 6655  (2000). 

\bibitem{Barrat-Berthier}
A. Barrat and L. Berthier, 
 {\it Real-space application of the mean-field description of spin-glass dynamics},
Phys. Rev. Lett. {\bf 87},  087204 (2001). 


\bibitem{Cudeku}
L. F. Cugliandolo, D. S. Dean, and J. Kurchan,
{\it Fluctuation-Dissipation theorems and entropy production 
in relaxational systems}, 
Phys. Rev. Lett. {\bf 79}, 2168 (1997). 

\bibitem{Behose}
L. Berthier, P. W. C. Holdsworth, and M. Sellitto, 
{\it Nonequilibrium Critical Dynamics of the 2D XY model}, 
 J. Phys. A {\bf 34}, 1805 (2001).

\bibitem{Ig}
J. L. Iguain, S. Bustingorry, and L. F. Cugliandolo, 
J. Stat. Mech.  P09008 (2007).   
J. L. Iguain, S. Bustingorry, A. B. Kolton, and L. F. Cugliandolo,
{\it Growing correlations and aging of an elastic line in a random potential},
Phys. Rev. B {\bf 80}, 094201 (2009).

\bibitem{Krzakala}
P. Calabrese, A. Gambassi, and F. Krzakala, 
{\it Critical aging of Ising ferromagnets relaxing from an ordered state},
J. Stat. Mech. P06016 (2006). 


\bibitem{fluctuation-theorem}
For reviews see, {\it e.g.}
G. Gallavotti, 
{\it Fluctuation Theorem and Chaos}
 Eur. Phys. J. B {\bf 64}, 315 (2008). 
F. Zamponi, 
{\it Is it possible to experimentally verify the fluctuation relation? 
A review of theoretical motivations and numerical evidence}, 
  J. Stat. Mech. P02008 (2007).
 J. Kurchan,
 {\it Six out of equilibrium lectures}
in Long-Range Interacting Systems, Les Houches Summer School 
Session 90, T. Dauxois, S. Ruffo, and L. F. Cugliandolo eds. 
(Oxford Univ. Press, 2009).
 
 \bibitem{FT-previous}
 M. Sellitto, 
 {\it Fluctuations of entropy production in driven glasses}, cond-mat/9809186.
 S. Sasa, {\it A fluctuation theorem for phase turbulence of chemical oscillatory waves}, 
nlin.CD/0010026.
 
\bibitem{Crri-FT}
A. Crisanti and F. Ritort, 
{\it Intermittency of glassy relaxation and the emergence of a non-equilibrium 
spontaneous measure in the aging regime}, 
Europhys. Lett. {\bf 66}, 253 (2004). 

 \bibitem{Bonetto}
 F. Zamponi, F. Bonetto, L. F. Cugliandolo, and J. Kurchan,
 {\it Fluctuation theorem for non-equilibrium relaxational systems driven by external forces},
J. Stat. Mech. P09013 (2005). 


\bibitem{Hayashi}
K. Hayashi and M. Takano,
{\it Temperature of a Hamiltonian system given as the effective temperature of a non-equilibrium steady state Langevin thermostat},
Phys. Rev. E {\bf 76}, 050104 (2007) and refs. therein. 


\bibitem{Mori}
A. Montanari and F. Ricci-Tersenghi, 
{\it Aging dynamics of heterogeneous spin models},
Phys. Rev. B {\bf 68}, 224429 (2003). 

\bibitem{Roma}
F. Rom\'a, S. Bustingorry, P. M. Gleiser, and D. Dom\'{\i}nguez,
{\it Strong Dynamical Heterogeneities in the Violation of the 
Fluctuation-Dissipation Theorem in Spin Glasses},
Phys. Rev. Lett. {\bf 98}, 097203 (2007). 

\bibitem{Chamon}
 C. Chamon, M. P. Kennett, H. E. Castillo, and L. F. Cugliandolo,
 {\it Separation of time-scales and reparametrization invariance for aging systems},
 Phys. Rev. Lett. {\bf 89}, 217201 (2002).

\bibitem{Chamon1}
H. E. Castillo, C. Chamon, L. F. Cugliandolo and M. P. Kennett, 
{\it Heterogeneous aging in spin glasses}, 
Phys. Rev. Lett. {\bf 88}, 237201 (2002). 
H. E. Castillo, C. Chamon, L. F. Cugliandolo, J. L.  Iguain, and M. P. Kennett, 
{\it Spatially heterogeneous ages in glassy dynamics}, 
Phys. Rev. B {\bf 68}, 134442 (2003). 

\bibitem{Chamon2}
C. Chamon, P.  Charbonneau, L. F. Cugliandolo, D.  R. Reichman and M. Sellitto, 
  {\it Out-of-equilibrium dynamic fluctuations in glassy systems}, 
Journal of Chemical Physics {\bf 121}, 10120 (2004).

 \bibitem{Chamon-Cugliandolo}
 C. Chamon and L. F. Cugliandolo, 
 {\it Fluctuations in glassy systems},
 J. Stat. Mech. P07022 (2007). 

\bibitem{Cukulepe}
L. F. Cugliandolo, J. Kurchan, P. Le Doussal, and L. Peliti,
{\it Glassy behaviour in disordered systems with non-relaxational dynamics}, 
 Phys. Rev. Lett. {\bf 78}, 350 (1997). 

\bibitem{Chcuyo} 
C. Chamon, L. F. Cugliandolo, and H. Yoshino,
{\it Fluctuations in the coarsening dynamics of the O(N) model with} $N to \infty$: 
{\it are they similar to those in glassy systems?},
J. Stat. Mech. P01006 (2006). 

\bibitem{Corberi11} C. Chamon, F. Corberi, and L. F. Cugliandolo,
{\it Fluctuations of two-time quantities and time-reparametrization invariance in spin-glasses},
in preparation.


 

\bibitem{Goldstein}
M. Goldstein,  {\it 
Viscous liquids and glass transition - a potential energy barrier picture},
 J. Chem. Phys. {\bf 51}, 3728 (1969).

\bibitem{Stillinger} F. H. Stillinger and T. A. Weber, 
{\it Hidden structure in liquids},
Phys. Rev. A {\bf 25}, 978 (1982). 

\bibitem{deDoyo} C. de Dominicis and A. P. Young, 
{\it Weighted averages and order parameters for the infinite range Ising spin glass}, 
J. Phys. A {\bf 6},  2063 (1983). 

\bibitem{Nieuwenhuizen}
T. M. Nieuwenhuizen,
{\it Thermodynamic picture of the glassy state gained from exactly solvable models},
Phys. Rev. E {\bf 61}, 267 (2000).

\bibitem{Biku}
G. Biroli and J. Kurchan,
{\it Metastable states in glassy systems},  
Phys. Rev. E {\bf 64}, 016101 (2001).

\bibitem{Bimo}
G. Biroli and R. Monasson,
{\it From inherent structures to pure states: Some simple remarks and examples}, 
Europhys. Lett. {\bf 50},  155 (2000).

\bibitem{Kob}
W. Kob, F. Sciortino, and P. Tartaglia, 
{\it Aging as dynamics in configurational space},
Europhys. Lett. {\bf 49}, 590 (2000). 


\bibitem{Dauchot} 
O. Dauchot, {\it Glassy behaviours in a-thermal
    systems, the case of granular media: a tentative review}, Lect.
  Notes in Phys. {\bf 716}, 161 (2007). H. A. Makse, J. Brujic, and
  S. F. Edwards, {\it Statistical Mechanics of Jammed Matter} in The
  physics of Granular Media (Wiley-VCH, 2004). A. Coniglio, A. Fierro, H. J. Hermann, and 
  M. Nicodemi (eds), {\it Unifying concepts in granular media and glasses}
  (Elsevier, Amsterdam, 2004). 
 
\bibitem{Jorge-Edwards}
 A. Barrat, J. Kurchan, V. Loreto, and M. Sellitto,
 {\it Edwards measures for powders and glasses},
 Phys. Rev. Lett. {\bf 85}, 5034 (2000). 
     {\it Edwards' measures: a thermodynamic construction for dense granular media and glasses},
    Phys. Rev. E {\bf 63}, 051301 (2001).
 
 \bibitem{Jorge-Hernan}
 H. Makse and J. Kurchan, 
 {\it Thermodynamic approach to dense granular matter: a numerical realization of a decisive 
experiment}, 
 Nature {\bf 415}, 614 (2002). 
F. Q. Potiguar and H. A. Makse
{\it Effective temperature and jamming transition in dense, gently sheared granular assemblies},
Eur. Phys. J. E {\bf 19}, 171 (2006).
 
 \bibitem{David-Alex} 
 D. S. Dean and A. Lef\`evre, {\it Steady state behaviour of mechanically perturbed 
 spin glasses and ferromagnets}, Phys. Rev. E {\bf 64}, 046110 (2001). 
 
 \bibitem{Brey} J. J. Brey, A. Prados and B. S\'anchez-Rey, 
 {\it Thermodynamic description in a simple model of granular compaction}, Physica A
 {\bf 275}, 310 (2000). 
 
 \bibitem{Godreche-Luck}
 C. Godr\`eche and J.-M. Luck,
 {\it Metastability in zero-temperature dynamics: Statistics of attractors},
J. Phys C. {\bf 17}, S2573 (2005).

\bibitem{Bertin-Dauchot}
E. Bertin, O. Dauchot, and M. Droz,
{\it Definition and relevance of nonequilibrium intensive thermodynamic parameters},
Phys. Rev. Lett. {\bf 96}, 120601 (2006).
E. Bertin, K. Martens, O. Dauchot, and M. Droz,
{\it Intensive thermodynamic parameters in nonequilibrium systems},
Phys. Rev. E {\bf 75}, 031120 (2007).


\bibitem{Pottier}
N. Pottier, {\it Aging properties of an anomalously diffusing particle},
Physica A {\bf 317}, 371 (2003). 

\bibitem{Ilg2} 
P. Ilg and J-L Barrat, 
{\it Effective temperatures in a simple model of non-equilibrium, non-Markovian dynamics}
in Statistical Physics of Ageing Phenomena and the Glass Transition,
{\bf 40}, 76 (2006).

\bibitem{OHern}
C. S. O'Hern, A. J. Liu, and S. R. Nagel, 
{\it Effective temperatures in driven systems: static vs. time-dependent
relations}, Phys. Rev. Lett. {\bf 93}, 165702 (2004). 


\bibitem{Henkel}
M. Henkel and M. Pleimling,
{\it Nonequilibrium Phase Transitions Volume 2 - Ageing and Dynamical Scaling 
far from Equilibrium}, 
Theoretical and Mathematical Physics (Springer, Heidelberg, 2010).

 \bibitem{Cocu} 
F. Corberi and L. F. Cugliandolo,  
{\it Fluctuations and effective temperatures in coarsening}, 
J. Stat. Mech.  P05010  (2009). 


 \bibitem{Godreche-Luck-critical}
 C. Godr\`eche and J.-M. Luck, 
{\it Response of non-equilibrium systems at criticality: 
Ferromagnetic models in dimension two and above},
J. Phys. A {\bf 33},  9141 (2000).

\bibitem{Julius}
J. Bonart, L. F. Cugliandolo, and A. Gambassi, 
{\it Critical Langevin dynamics of the O(N) Ginzburg-Landau 
model with correlated noise}, in preparation.

\bibitem{Annibale}
A. Annibale and P. Sollich,
{\it Dynamic heterogeneities in critical coarsening: 
Exact results for correlation and response fluctuations in 
finite-sized spherical models}, 
J. Stat. Mech. P02064 (2009). 

\bibitem{Joubaud}
S. Joubaud, B. Percier, A. Petrosyan A, and S. Ciliberto,
{\it Aging and Effective Temperatures Near a Critical Point}, 
 Phys. Rev. Lett. {\bf 102}, 130601 (2009).

\bibitem{Mayer-etal}
P. Mayer, L. Berthier, J. P. Garrahan, and P. Sollich, 
{\it Fluctuation-dissipation relations in the non-equilibrium critical dynamics of Ising models}, 
Phys. Rev. E {\bf 68}, 016116 (2003).


\bibitem{1dIC}
C. Godr\`eche and J.-M. Luck,
{\it Response of non-equilibrium systems at criticality: Exact results for the Glauber-Ising chain},
J. Phys. A {\bf 33},  1151 (2000).
E. Lippiello and M. Zannetti, 
{\it Fluctuation dissipation ratio in the one dimensional kinetic Ising model},
 Phys. Rev. E {\bf 61}, 3369 (2000).


\bibitem{Kob-Barrat}
W. Kob and J. L. Barrat, 
{\it Fluctuations, response and aging dynamics in a simple glass-forming liquid out of equilibrium},
Eur. Phys. J. B {\bf 13}, 319 (2000).

\bibitem{DiLeonardo}
R. Di Leonardo, L. Angelani, G. Parisi, and G. Ruocco,
{\it Off-Equilibrium effective temperature in monatomic Lennard-Jones glass},
Phys. Rev. Lett. {\bf 84}, 6054 (2000). 
 N. Gnan, C., T. B. Schroeder, and J. C. Dyre,
{\it Predicting the effective temperaure of a glass},
Phys. Rev. Lett. {\bf 104}, 125902 (2010). 

\bibitem{Grigera03}
T. S. Grigera, V. Martin-Mayor, G. Parisi, and P. Verrocchio,
{\it Asymptotic aging in structural glasses},
Phys. Rev. B {\bf 70}, 014202 (2004).

\bibitem{Berthier}
L. Berthier, {\it Efficient measurement of linear susceptibilities in molecular simulations: Application to aging supercooled liquids}, 
Phys. Rev. Lett. {\bf 98}, 220601 (2007). 

\bibitem{Gradenigo}
G. Gradenigo, A.  Sarracino, D. Villamaina, T. S. Grigera, 
and A. Puglisi, 
{\it The Ratchet effect in an ageing glass}, 
J. Stat. Mech. L12002 (2010). 


\bibitem{kinetically-constrained}
F. Ritort and P. Sollich,
{\it Glassy dynamics of kinetically constrained models},
Adv. in Physics, {\bf 52}, 219 (2003). 
J. P. Garrahan, P. Sollich, and C. Toninelli,
{\it Kinetically Constrained Models},
arXiv:1009.6113 in~\cite{book-lengths}.

\bibitem{Leonard}
S. Leonard, P. Mayer, P. Sollich, L. Berthier, and J. P. Garrahan,
{\it Non-equilibrium dynamics of spin facilitated glass models},
 J. Stat. Mech. P07017 (2007).


\bibitem{Nathan1}
T. Grigera and N. E. Israeloff, 
{\it Observation of fluctuation-dissipation-theorem violations in a structural glass}, 
 Phys. Rev. Lett. {\bf 83}, 5038 (1999).

\bibitem{Hernan2}
P. Wang, C. Song, and H. A. Makse,
{\it Dynamic particle tracking reveals the aging temperature of a colloidal glass},
 Nature Physics {\bf 2}, 526 (2006). 

\bibitem{Ciliberto}
L. Buisson, L. Bellon, and S. Ciliberto,
{\it Intermittency in aging},
J. Phys.: Condens. Matt. {\bf 15}, S1163 (2003). 

\bibitem{Strachan} 
D. R. Strachan, G. C. Kalur, and  S. R. Raghavan, 
  {\it Size-dependent diffusion in an aging colloidal glass},
    Phys. Rev. E {\bf 73}, 041509 (2006).

\bibitem{Abou}
B. Abou and F. Gallet,
{\it Probing a nonequilibrium Einstein Relation in an Aging Colloidal Glass},
 Phys. Rev. Lett. {\bf 93}, 160603 (2004).
 B. Abou, F. Gallet, P. Monceau, and N. Pottier, 
{\it Generalized Einstein Relation in an aging colloidal glass},
Physica A {\bf 387}, 3410 (2008).

\bibitem{Greinert}
N. Greinert, T. Wood, and P. Bartlett,
{\it Measurement of Effective Temperatures in an Aging Colloidal Glass},
 Phys. Rev. Lett. {\bf 97}, 265702 (2006).

\bibitem{Jabbari} S. Jabbari-Farouji, D. Mizuno, M. Atakhorrami, F. C.
  MacKintosh, C. F. Schmidt, E. Eiser, G. Wegdam, and D. Bonn, {\it
    Fluctuation-dissipation theorem in an aging colloidal glass},
  Phys. Rev. Lett. {\bf 98}, 108302 (2007).

 \bibitem{Jop}
P. Jop, J. Ruben G\'omez-Solano, A. Petrosyan, and S. Ciliberto, 
{\it Experimental study of out of equilibrium fluctuations in a 
colloidal suspension of Laponite using optical traps},
J. Stat. Mech.  P04012 (2009).

\bibitem{Maggi}
C. Maggi, R. Di Leonardo, J. C. Dyre, and G. Ruocco,
{\it Generalized fluctuation-dissipation relation and 
effective temperature in off-equilibrium colloids}, 
Phys. Rev. B {\bf 81}, 104201 (2010). 

\bibitem{Nathan2} H. Oukris and N. E. Israeloff, {\it Nanoscale nonequilibrium
dynamics and the fluctuation dissipation relation in an aging polymer glass},
Nature Physics {\bf 6}, 135 (2010).


\bibitem{Fisher-Huse} 
D. S. Fisher and D. A. Huse, 
{\it Ordered Phase of Short-Range Ising Spin-Glasses}, 
Phys. Rev. Lett. 56, 1601 (1986);
{\it Equilibrium behavior of the spin-glass ordered phase}, 
Phys. Rev. B {\bf 38}, 386 (1988);
{\it Nonequilibrium dynamics of spin glasses},
Phys. Rev. B {\bf 38}, 373 (1988).

\bibitem{Kawamura-model}
H. Kawamura, 
{\it The ordering of XY spin glasses},
arXiv:1102.3496, and refs. therein.

\bibitem{trap}
J. P. Bouchaud, 
{\it Weak ergodicity breaking and aging in disordered systems},
J. Phys. (France) I {\bf 2}, 1705 (1992).

\bibitem{Juan}
E. Marinari, G. Parisi, F. Ricci-Tersenghi, J. Ruiz-Lorenzo, and F. Zuliani,
{\it Replica symmetry breaking in short range spin glasses: 
A review of the theoretical foundations and of the numerical evidence}, 
 J. Stat. Phys. {\bf 98}, 973 (2000).

\bibitem{Kawamura-FDT} 
H. Kawamura,
{\it Fluctuation-dissipation ratio of the Heisenberg spin-glass},
 Phys. Rev. Lett. {\bf 90}, 237201 (2003).
 
\bibitem{Grempel}
L. F. Cugliandolo, D. R. Grempel, J. Kurchan, and E. Vincent, 
{\it A search for fluctuation-dissipation theorem violations in spin-glasses from 
susceptibility data}, 
 Europhys. Lett. {\bf 48}, 699 (1997). 

\bibitem{Ocio}
D. H\'erisson and M. Ocio, 
{\it Fluctuation-dissipation ratio of a spin glass in the aging regime},
Phys. Rev. Lett. {\bf 88}, 257202 (2002). 
{\it Off-equilibrium fluctuation-dissipation relation in a spin glass},
Eur. Phys. J. B {\bf 40}, 283 (2004).

\bibitem{Denis-Lhote} D. L'H\^ote, S. Nakamae, F. Ladieu, V. Mosser,
  A. Kerlain, and M. Konczykowski, 
  {\it A local noise measurement device for magnetic physical systems}, 
  J. Stat. Mech. P01027 (2009).

\bibitem{Denis-Lhote2} K. Komatsu, D. L'H\^ote, S. Nakamae, V. Mosser,
  M. Konczykowski, E. Dubois, V. Dupuis, R. Perzynski, {\it
    Experimental evidence of fluctuation-dissipation theorem violation
    in a superspin glass}, arXiv:1010.4012.

\bibitem{spin-ice} 
 C. Nisoli, J. Li, X. Ke, D. Garand, P. Schiffer, and V. H. Crespi, 
{\it Effective Temperature in an Interacting Vertex
  System: Theory and Experiment on Artificial Spin Ice}, 
Phys. Rev. Lett. {\bf 105}, 047205 (2010).


\bibitem{Berthier-Barrat}
 L. Berthier and  J.-L. Barrat,
{\it Nonequilibrium dynamics and fluctuation-dissipation relation in a sheared fluid},
 J. Chem. Phys. {\bf 116}, 6228 (2002). 

\bibitem{Bebaku}
 L. Berthier, J-L Barrat, and J. Kurchan, 
{\it Two-time scales, two-temperature scenario for nonlinear rheology}, 
Phys. Rev. E {\bf 61}, 5464-5472 (2000). 

\bibitem{Angelani05}
F. Zamponi, G. Ruocco, and L. Angelani,
{\it Generalized fluctuation relation and effective temperatures in a 
driven fluid}, 
Phys. Rev. E {\bf 71}, 020101(R) (2005).

\bibitem{Kolton1}
A. B. Kolton, R. Exartier, L. F. Cugliandolo,
 D. Dominguez, and N. Gr\/onbech-Jensen
{\it Effective temperature in driven vortex lattices with random pinning},
Phys. Rev. Lett. {\bf 89}, 227001 (2002).


\bibitem{Danna}
G. D'Anna, P. Mayor, A. Barrat, V. Loreto, and F. Nori, 
{\it Observing Brownian motion in vibration-fluidized granular matter}, 
 Nature {\bf 424}, 909 (2003).

\bibitem{Wang}
 P. Wang, C. Song, C. Briscoe, and H. A. Makse,
{\it Particle dynamics and effective temperature of jammed granular matter 
in a slowly sheared 3D Couette cell},
 Phys. Rev. E {\bf 77}, 6 (2008).
C. Song, P. Wang, and H. A. Makse, 
{\it Experimental measurement of an effective temperature for
jammed granular materials}, Proc. Nat. Acad. Sci. {\bf 102}, 2299 (2005).
 
\bibitem{Dennin} 
M. Twardos and M. Dennin, 
{\it Asymmetric response of a jammed plastic bead raft}, 
Phys. Rev. Lett. {\bf 97}, 110601 (2006).

\bibitem{Kolton}
A. B. Kolton, D. Dom\'{\i}nguez, and N. Gr\/onbech-Jensen,
{\it Hall noise and transverse freezing in driven vortex lattices},
Phys. Rev. Lett. {\bf 83}, 3061 (1999). 

\bibitem{Behringer}
J. F. Geng and R. P. Behringer, 
{\it Diffusion and mobility in a stirred dense granular material}, 
Phys. Rev. Lett. {\bf 93}, 238002 (2004).


\bibitem{Ilg1}
P. Ilg and J-L Barrat, 
{\it Driven activation vs. thermal activation}, EPL {\bf 79}, 26001 (2007).

\bibitem{Haxton} 
T. K. Haxton and A. J. Liu, 	
{\it Activated dynamics and effective temperature in a steady state sheared glass},
Phys. Rev. Lett. {\bf 99}, 195701 (2007).

\bibitem{Abate}
A. R. Abate and D. J. Durian, 
{\it Effective temperatures and activated dynamics for a two-dimensional
 air-driven granular system on two approaches to jamming}, 
Phys. Rev. Lett. {\bf 101}, 245701 (2008).
 

 \bibitem{Lu}
 T. Lu, J. Hasty, and P. G. Wolynes, 
{\it Effective temperature in stochastic kinetics and gene networks}, 
Biophys. J. {\bf 91}, 84 (2006).
 
\bibitem{Martin} 
 P. Martin, A. J. Hudspeth, and F. J\"ulicher, 
 {\it Comparison of a hair bundle's spontaneous oscillations with its response to mechanical 
 stimulation reveals the underlying active process}, 
  Proc. Nac. Acad. Sc. USA {\bf 98}, 14380
(2001).

\bibitem{mizuno07}
D. Mizuno, C. Tardin, C.~F. Schmidt, and F.~C. MacKintosh,
{\it Nonequilibrium mechanics of active cytoskeletal networks}, 
Science {\bf 315}, 370 (2007).

\bibitem{Morozov}
K. I. Morozov and L. M. Pismen, 
{\it Motor-driven effective temperature and viscoelastic response of active matter}, 
Phys. Rev. E {\bf 81},  061922 (2010). 

\bibitem{Jack}
R. L. Jack, M. F. Hagan, and D. Chandler,
{\it Fluctuation-dissipation ratios in the dynamics of self-assembly}, 
Phys Rev E {\bf 76}, 021119 (2007). 

\bibitem{Yair}
E. Ben-Isaac, YK Park, G. Popescu, F. L. H. Brown, N. S. Gov, and Y. Shokef, 
{\it Effective temperature of red blood cell membrane fluctuations}, 
arXiv:1102.4508.

 \bibitem{Harada-Sasa} 
T. Harada and S-I Sasa, 
 {\it Fluctuations, responses and energetics of molecular motors}
Mathematical biosciences {\bf 207}, 365 (2007), and refs. therein.

\bibitem{Kolton-ratchet}
A. B. Kolton, 
{\it Transverse rectification of disorder-induced fluctuations in a driven system},
Phys. Rev. B {\bf 75},  020201 (2007). 

\bibitem{Shen} 
T. Shen and P. G. Wolynes, 
{\it Stability and dynamics of crystals and glasses of motorized particles}, 
Proc. Nac. Acad. Sc. USA
{\bf 101}, 8547 (2004); 
{\it Nonequilibrium statistical mechanical models for cytoskeletal assembly: 
Towards understanding tensegrity in cells}, 
Phys. Rev. E {\bf 72}, 041927 (2005).

\bibitem{Loi}
D. Loi, S. Mossa, and L. F. Cugliandolo, 
 {\it Effective temperature of active matter},
Phys. Rev. E {\bf 77}, 051111 (2008); 
{\it Effective temperature of active complex matter}, 
Soft Matter {\bf 7}, 3726 (2011).

\bibitem{palacci10}
J. Palacci, C. Cottin-Bizonne, C. Ybert, and L. Bocquet,
{\it Sedimentation and effective temperature of active colloidal suspensions},
Phys. Rev. Lett. {\bf 105}, 088304 (2010).
%

\bibitem{Tailleur}
J. Tailleur and M. E. Cates, 
{\it Sedimentation, trapping, and rectification of dilute bacteria}, 
EPL {\bf 86}, 60002 (2009).

\bibitem{joly10}
L. Joly, S. Merabia, and J.-L. Barrat, 
{\it Effective temperatures of a heated Brownian particle}, 
 arXiv:1101.2758.
%

%


\bibitem{Langer} J. S. Langer, 
{\it Dynamics of shear-transformation zones in amorphous plasticity: 
Formulation in terms of an effective disorder temperature},
Phys. Rev. E {\bf 70}, 041502 (2004).
  E. Bouchbinder and J. S. Langer, {\it Nonequilibrium thermodynamics
    of driven amorphous materials. II. Effective-temperature theory},
  Phys. Rev. E {\bf 80}, 031132 (2009).

\bibitem{Shi}
Y. Shi, M. B. Katz, H. Li, and M. L. Falk, 
{\it Evaluation of the Disorder Temperature 
and Free-Volume Formalisms via Simulations of Shear Banding in Amorphous Solids}, 
Phys. Rev. Lett. {\bf 98}, 185505 (2007). 

\bibitem{Yoshino10}
H. Yoshino and M. M\'ezard, 
{\it 
Emergence of rigidity at the structural glass transition: A first principle computation}, 
Phys. Rev. Lett. {\bf 105},  015504 (2010).

\bibitem{Yoshino}
H. Yoshino, in preparation.


\bibitem{Naert}
V. Grenard, N. Garnier, and A. Naert,
{\it Fluctuation-dissipation relation on a Melde string in a turbulent flow, considerations on a 
"dynamical temperature"}, 
J. Stat. Mech. L09003 (2008).

\bibitem{Daviaud}
R. Monchaux, P-P Cortet, P-H Chavanis, A. Chiffaudel, 
F. Daviaud, P. Diribarne, and B. Dubrulle, 
{\it Fluctuation-Dissipation Relations and statistical temperatures in a 
turbulent von K\'arm\'an flow},
Phys. Rev. Lett. {\bf 101}, 174502 (2008).



\bibitem{Arbicu2}
C. Aron, G. Biroli, and L. F. Cugliandolo, 
{\it Driven Quantum Coarsening}, Phys. Rev. Lett. {\bf 102}, 050404 (2009);
{\it Coarsening of disordered quantum rotors under a bias voltage}, 
Phys. Rev. B {\bf 82}, 174203 (2010).

\bibitem{static-quantum}
L. F. Cugliandolo, D. R. Grempel, G. S. Lozano, H. Lozza, and C. A. da Silva Santos,
{\it Dissipative effects on quantum glassy systems},
 Phys. Rev. B {\bf 66}, 014444 (2002). 
G. Biroli and L. F. Cugliandolo, 
{\it Quantum TAP equations},
Phys. Rev. B {\bf 64}, 014206 (2001).

\bibitem{Lili1}
L. Arrachea and L. F. Cugliandolo, 
{\it Study of a fluctuation-dissipation relation of a dissipative driven mesoscopic system}, 
Europhys. Lett. {\bf 70}, 642Ð648 (2005).

\bibitem{Lili2}
A. Caso, L. Arrachea, and G. Lozano, 
{\it Local temperatures and heat flow in quantum driven systems}, 
Phys. Rev. B {\bf 81}, 041301 (2010); 
{\it Relation between local temperature gradients and the direction 
of heat flow in quantum driven systems}, arXiv:1102.4491.

\bibitem{Rigol}
M. Rigol, V. Dunjko, and M. Olshanii, 
{\it Thermalization and its mechanism for generic isolated quantum systems},
Nature {\bf 452}, 854 (2008). 

\bibitem{Rossini}
 D. Rossini, A. Silva, G. Mussardo, G. E. Santoro,
{\it Effective thermal dynamics following a quantum
  quench in a spin chain Authors},
 Phys. Rev. Lett. {\bf 102}, 127204
  (2009). D. Rossini, S. Suzuki, G. Mussardo, G. E. Santoro, and A. Silva,
 {\it Long time dynamics following a quench in an
  integrable quantum spin chain: local versus non-local operators and
  effective thermal behavior}, Phys. Rev. B {\bf 82}, 144302 (2010). 

\bibitem{Foini}
L. Foini, L. F. Cugliandolo, and A. Gambassi, 
in preparation. 

\bibitem{Racz}
V. Eisler, O. Legeza, Z. Racz,
{\it Fluctuations in subsystems of the zero temperature XX chain: 
Emergence of an effective temperature}, 
J. Stat. Mech. P11013 (2006). 


\end{thebibliography}
\end{document}